\documentclass[notitlepage,prd,reprint,superscriptaddress,twocolumn]{revtex4-1}
\usepackage{graphicx,amsmath,bm,natbib}

\newcommand{\Sigc}{\ensuremath{\Sigma_{\rm crit}}}
\renewcommand{\epsilon}{\varepsilon}

\begin{document}

\title{Three-point galaxy-galaxy lensing as a probe of dark matter halo shapes}
\author{Susmita Adhikari}
\affiliation{Department of Astronomy, University of Illinois at Urbana-Champaign}
\author{Chun Yin Ricky Chue}
\affiliation{Department of Astronomy, University of Illinois at Urbana-Champaign}
\author{Neal Dalal}
\affiliation{Department of Astronomy, University of Illinois at Urbana-Champaign}
\affiliation{Department of Physics, University of Illinois at Urbana-Champaign}

\begin{abstract}
We propose a method to measure the ellipticities of dark matter halos using the lens-shear-shear 3-point correlation function.  This method is immune to effects of galaxy-halo misalignments that can potentially limit 2-point galaxy-galaxy lensing measurements of halo anisotropy.  Using a simple model for the projected mass distributions of dark matter halos, we construct an ellipticity estimator that sums over all possible triangular configurations of the 3-point function. By applying our estimator to halos from N-body simulations, we find that systematic errors in the recovered ellipticity will be at the $\lesssim 5\%$ fractional level.  We estimate that future imaging surveys like LSST will have sufficient statistics to detect halo ellipticities using 3-point lensing.
\end{abstract}

\maketitle

\section{Introduction}
\label{intro}

In the cold dark matter model of cosmological structure formation, galaxies are believed to form inside of virialized objects called dark matter halos.  The properties of these halos, like their internal structure or abundance, are related to the background cosmology and to the physics of dark matter particles.  One example of this is the ellipticity of dark matter halos.   In CDM cosmologies, halos are found to be triaxial, with axis ratios of the order of 0.5:1, with a significant scatter from object to object \citep{Dubinski1991,Jing2002,Allgood2006,Schneider2012}.  Alternative models, like self-interacting dark matter (SIDM) can produce significantly different shapes.  Pure SIDM simulations generally produce halos with rounder shapes than CDM simulations \citep{Dave2001, Peter13, Rocha13}, although the effects of baryons can  modify these results \citep{2013arXiv1311.6524K}.

Therefore, measurements of the shapes of dark matter halos may be used to probe the nature of dark matter.  Accordingly, multiple groups have attempted to measure halo shapes using a variety of probes.  In our own Galaxy, several groups have attempted to model the dynamics of the Sagittarius tidal
stream in order to infer the underlying shape of the Milky Way's halo \citep{Helmi2004,Law2010,Deg2013,Ibata2013,VeraCiro2013}.  In other galaxies, halo shapes have been probed using strong lensing and stellar dynamics \citep{vandeVen2010,Suyu2012} on small scales, and satellite dynamics on larger scales \citep{Bailin2008}.  

Another probe of dark matter halo properties is weak gravitational lensing.  The average radial profiles of dark matter halos have been inferred with high precision through measurement of the two-point cross-correlation between galaxies and tangential shear, called galaxy-galaxy lensing \citep{Sheldon2004,Mandelbaum06,Johnston2007}.  Circularly averaged statistics are insensitive to halo ellipticity, but in principle, anisotropy could be constrained by measuring shear not only as a function of radius $r$, but position angle $\theta$ as well.  Unfortunately, because dark matter halos are dark, we cannot determine the orientations of halos, making it impossible to measure shear profiles as a function of position angle relative to the halo principal axes.  We can, however, measure shear as a function of the position angle relative to the lens galaxies' principal axes.  If halos are perfectly aligned with their central galaxies, then such measurements may be used to determine the average halo ellipticity.  This is the approach that has been used by most previous work \citep{Hoekstra2004,Mandelbaum2006,Parker2007,Uitert}.  This previous work, however, has yielded inconclusive results.  For example, \citet{Uitert} report an average projected ellipticity of $e=0.38\pm0.26$, which is consistent both with CDM predictions and with completely isotropic halos.  Currently, statistical errors are a principal limitation of this measurement, but with the vastly increased sample sizes provided by future imaging surveys like LSST, the statistical errors may be reduced sufficiently to detect the expected signal.   More worryingly, this method is likely limited by potentially severe systematic effects.  First, the assumption that galaxies and their halos are perfectly aligned may be unrealistic \citep{Tenneti2014}.  \citet{Bett2012} has argued that significant misalignments between galaxies and halos may be quite typical; the median misalignment angle in their simulations was $\sim 38^\circ$.  Random misalignments act to wash out the halo anisotropy signal from galaxy-galaxy lensing.  Even worse, they complicate the interpretation of any measured anisotropy signal.  Without knowledge of the misalignment distribution, we will not know how to translate stacked lensing signals into constraints on halo axis ratios.  This effect is also not the only possible systematic.  For example, if lens galaxies and background source galaxies are both lensed by foreground structures, this common lensing will tend to align their observed shapes, thereby contaminating the halo anisotropy signal \citep{Howell2010}.  Because of these systematic limitations, an alternative approach for measuring halo shapes with galaxy-galaxy lensing may be required -- ideally, a method that does not require galaxies to align with their host halos.  Such an approach is suggested by the recent work of \citet{Simon2012}, who find that halo ellipticities affect galaxy-galaxy lensing 3-point correlation functions.  Although most previous work on galaxy-galaxy lensing has focused on 2-point statistics, higher order correlation functions are now becoming measurable in modern imaging surveys \citep{Simon2013,Fu2014}.  In this paper, we explore how halo ellipticities may be determined from measurements of the galaxy-shear-shear 3-point function.

\section{Mass model}
\label{sec:model}

\begin{figure*}
\centering
\includegraphics[width=0.9\textwidth,trim=0in 0.5in 0in 0.5in, clip]{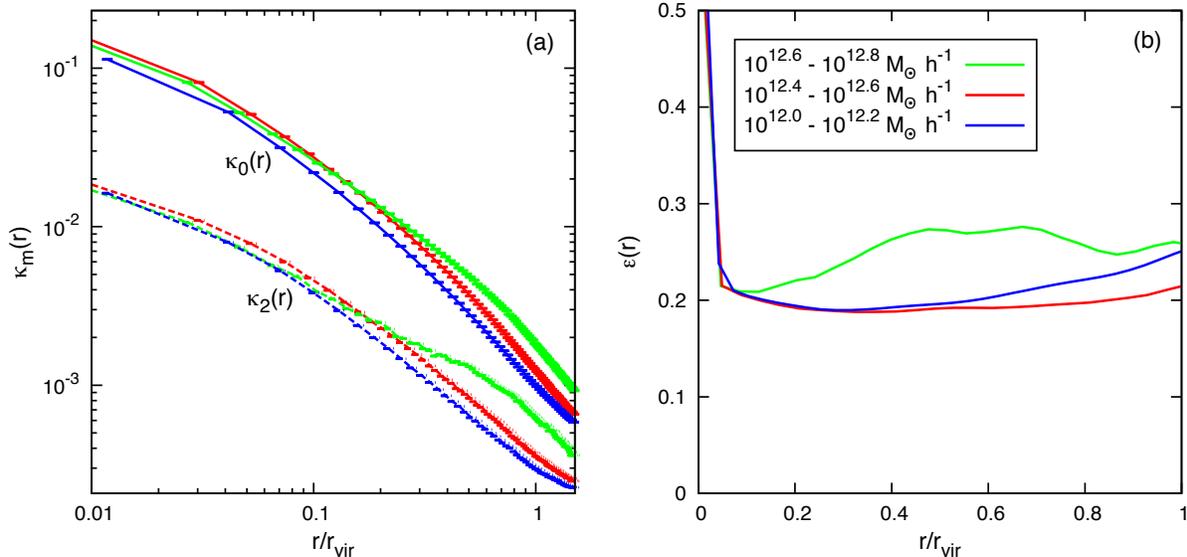}
\caption{(a): Plot of the multipole moments of stacked halos. The solid curves show the isotropic component (monopole, $\kappa_0$) of the surface density profile and the dashed curves show the $\cos 2\theta$ component (quadrupole, $\kappa_2$). (b): Radial dependence of ellipticity, which we define as $\epsilon(r)\equiv\kappa_2/(\eta\,\kappa_0)$, for three different mass bins. The blue, red and green colors correspond to three different mass bins.  Note that, although the multipole moments vary by orders of magnitude, the ellipticity remains nearly constant across much of the range of interest.}
\label{fig:ratio_1}
\end{figure*}

The 3D density profiles of halos in dissipationless CDM simulations have axis ratios of order $q\approx 0.5$, slowly varying with radius \citep{Jing2002,Allgood2006}.  Similarly, the 2D projected surface density $\Sigma$ is anisotropic, with axis ratios closer to $q\sim 0.7$, again slowly increasing with radius.  Because $q$ is nearly constant with radius, we can write $\Sigma\propto R^{-\eta}$, where $R=(x^2 + y^2/q^2)^{1/2}$ is an ellipsoidal radial coordinate, and $\eta$ is the logarithmic slope of the projected surface density.  We will find it convenient below to work with the multipole moments of the density profile.  In the limit of small ellipticity, we can write the multipole expansion of $\Sigma$ in terms of $q$,
\begin{eqnarray}
\Sigma(r,\theta) & \propto & r^{-\eta} \left[1 + \epsilon\, \eta \cos 2\theta + {\cal O}(\epsilon^2)\right]\nonumber \\ 
 &\equiv& \Sigma_0(r) + \Sigma_2(r) \cos 2\theta + \ldots 
\end{eqnarray}
where the multipole $\Sigma_m(r)$ is the coefficient of the $e^{i m\theta}$ component of the azimuthal behavior, and we use $\epsilon=(1-q^2)/\left[2(1+q^2)\right]$ to parameterize the ellipticity.  For the typical axis ratios found in simulated halos, $\epsilon \sim 0.2$, so we neglect higher order terms in the expansion.  

We therefore model the mass distributions of halos as the sum of a monopole and quadrupole, and further assume that 
\begin{equation}
\epsilon \approx \frac{\Sigma_2(r)}{\eta(r) \Sigma_0(r)}
\label{eq:epsilon}
\end{equation}
where $\eta=d\log\Sigma_0/d\log r$.  In Figure \ref{fig:ratio_1}, we plot $\epsilon$ as defined in Eqn.\ \eqref{eq:epsilon}, measured from stacked profiles of projected halos taken from the Bolshoi simulation \citep{Klypin2011}.  We measure multipole moments from the particle positions, using
\begin{equation}
\Sigma_m(r) = \sum_i \frac{m_p \delta(r-r_i)\,e^{i m \theta_i}}{2\pi r_i},
\end{equation}
where $m_p$ is the particle mass, and $r_i$ and $\theta_i$ are the radius and azimuthal angle for particle $i$.  After computing $\Sigma_0(r)$ and $\Sigma_2(r)$ for each halo, we then stack the halos to compute $\langle\Sigma_0\rangle(r)$ and $\langle|\Sigma_2|\rangle(r)$, and then $\epsilon$.  As expected, the ellipticity is fairly constant with radius, except very near the halo center where $\eta = d\log\Sigma_0/d\log r \to 0$.  
Because $\epsilon$ is nearly constant with radius, then the radial dependence of the quadrupole may be predicted from the monopole, whose mean $\langle\Sigma_0(r)\rangle$ may be determined from real galaxy halos using galaxy-galaxy lensing 2-point statistics.  

Specifically, the mean tangential shear $\langle\gamma_+\rangle$ profile around halos is related to the mean monopole via \citep{bartelmann2001weak}
\begin{equation}\label{eq:tangential_shear}
 \langle\gamma_+\rangle=\frac{\Delta\Sigma(r)}{\Sigc}=\frac{\bar\Sigma_0(<r)-\Sigma_0(r)}{\Sigc},
\end{equation}
where the lensing critical density $\Sigc$ is defined as
\begin{equation}
 \Sigc=\frac{c^2}{4\pi G}\frac{D_s}{D_{ds}D_d},
\end{equation} 
and  $D_d$ and $D_s$ are the angular diameter distances to the lens and the source respectively, while $D_{ds}$ is the angular diameter distance from the lens to the source.

For circularly symmetric lenses, the tangential shear is the only nonzero component of the shear.  When the surface density is anisotropic, however, the other component ($\gamma_\times$) becomes nonzero.  In the same way that we can decompose the surface density into angular multipoles $\Sigma_m(r)$, we can similarly decompose the shear into multipoles $\bm{\gamma}^{(m)}(r)$.  The relation between the density and shear multipoles is straightforward.  For convenience, we follow conventional notation and define the convergence as $\kappa=\Sigma/\Sigc$, and define a 2D lensing potential $\psi$ via 
\begin{equation}
\nabla^2\psi = 2\kappa
\end{equation}
where the gradient is with respect to sky coordinates.  In polar coordinates, this equation becomes
\begin{equation}\label{eq:kappa_psi}
\kappa(r,\theta)=\frac{1}{2}\left[\frac{\partial^2}{\partial r^2} + \frac{1}{r}\frac{\partial}{\partial r} + \frac{1}{r^2}\frac{\partial^2}{\partial\theta^2}\right]\psi.
\end{equation}
The two components of the shear are given by
\begin{eqnarray}\label{eq:tan}
\gamma_+ &=& \left[-\frac{\partial^2}{\partial r^2} + \frac{1}{r}\frac{\partial}{\partial r} + \frac{1}{r^2}\frac{\partial^2}{\partial\theta^2}\right]\psi\\
\label{eq:cross}
\gamma_{\times} &=& \left[-\frac{2}{r}\frac{\partial^2}{\partial r\partial \theta} + \frac{2}{r^2} \frac{\partial}{\partial\theta}\right]\psi
\end{eqnarray}

Next, let us decompose these fields into angular multipoles 
\begin{eqnarray} \label{eq:mult_exp}
\psi\left(r, \theta\right) &=& \sum_{m = -\infty}^{\infty} \psi_m(r)e^{im\theta}\\
\kappa\left(r, \theta\right) &=& \sum_{m = -\infty}^{\infty} \kappa_m(r)e^{im\theta}\nonumber.
\end{eqnarray}
Explicitly, 
\begin{eqnarray}
m = 0:\quad\kappa_0(r) &=&\frac{1}{2\pi}\int^{2\pi}_{0} \kappa(r,\theta)d\theta \\
m\geq 1:\quad\kappa_m(r) &=&\frac{1}{\pi}\int^{2\pi}_{0} \kappa(r,\theta) \cos m\theta\, d\theta.\nonumber
\end{eqnarray}
Solving the 2-d Poisson equation \eqref{eq:kappa_psi}, we obtain the multipole moments of $\psi$, 
\begin{eqnarray}\label{eq:multipole}
\psi_0(r) &=& \ln r\int_0^r r' \kappa_0(r') dr' + \int_r^{\infty} r' \ln r' \kappa_0(r') dr' \nonumber\\
\psi_m(r) &=& -\frac{1}{2m}\left[r^{-m}\int_0^r r'^{(m + 1)} \kappa_m(r') dr' \right. \nonumber\\ &&\qquad\left.+~ r^m\int_r^{\infty} r'^{(1-m)}\kappa_m(r') dr'\right].
\end{eqnarray}
Then, using Eqns.\ (\ref{eq:tan}) and (\ref{eq:cross}), we may obtain the multipole moments of the two shear components.  Because we keep only $m=0$ and $m=2$, and because we assume that $\kappa_2(r) = \epsilon\,\eta(r)\,\kappa_0(r)$, we have 
\begin{eqnarray}\label{eq:g_shear}
\gamma^{(0)}_+(r) &=& \frac{2}{r^2}\int_0^r r'\kappa_0(r')dr' - \kappa_0(r) \\
g_+(r) &=& \left[-\kappa_0(r)\eta(r) + \frac{3}{r^{4}}\int_0^r r'^{3}\kappa_0(r')\eta(r') dr'\right. \nonumber \\ && \qquad
\left.+ \int_r^{\infty}\frac{\kappa_0(r')\eta(r')}{r'} dr'\right] \\
g_\times(r) &=& \left[\frac{3}{r^{4}}\int_0^r r'^{3}\kappa_0(r')\eta(r') dr' \right. \nonumber \\ && \qquad
\left.-\int_r^{\infty}\frac{\kappa_0(r')\eta(r')}{r'} dr'\right],
\end{eqnarray}
where we have defined, for the purpose of convenience, the functions $g_+$ and $g_\times$ such that the quadrupole components of the shear are $\gamma_+^{(2)} = \epsilon\,g_+(r)\cos 2\theta$ and $\gamma^{(2)}_\times = \epsilon\,g_\times(r)\sin 2\theta$.  Note that, by definition, $\gamma^{(0)}_\times=0$.

Given this model for the mass distributions of lenses, we can predict the shear at all locations around the lenses.  The one unknown parameter is the ellipticity $\epsilon$, which defines the amplitude of the quadrupole moment $\kappa_2$ in terms of the (known) monopole moment $\kappa_0$.  Because we have an expression for the shear at all locations, we can construct an estimator for the quantity $\epsilon$.

\subsection{Three-Point Estimator}

As discussed in \S\ref{intro}, \citet{Simon2012} have shown that lensing 3-point functions are sensitive to halo ellipticities.  However, they also show that lensing 3-point functions are also sensitive to many other terms, making it difficult to disentangle the signal in the bispectrum generated by halo ellipticity.  Fortunately, 
given our model for halo mass distributions, it is straightforward for us to construct an estimator to measure halo ellipticity from lensing correlation functions.  Following \citep{Simon2012}, we focus on the lens-shear-shear 3-point function.  Measurement of this correlation function involves stacking the shear measured from pairs of source galaxies behind foreground lens galaxies.  Because the number density of source pairs is low, especially at the small radii of interest for measuring internal halo properties ($r<r_{\rm vir}$), we assume that shape noise in the source galaxies dominates  measurement uncertainties.  That is, we neglect the signal covariance compared to Poisson fluctuations in source counts.  Because Poisson noise is white noise, the optimal estimator is then simply proportional to the expected signal from our model.  

\begin{figure}
\centering
\includegraphics[width=0.4\textwidth]{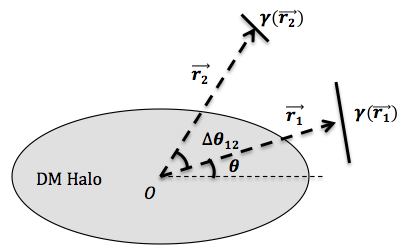}
\caption{Lens-shear-shear three point correlation function. We correlate the shear at sky positions $\vec r_1$ and $\vec r_2$ relative to foreground lens galaxies, and construct an estimator summing all such triangular configurations in the sky.}
\label{fig:three_pt_corr}
\end{figure}

 We therefore estimate the average lens ellipticity by summing over all lens-source-source triangles, weighting each triangle with a filter {\bf F} that is given by the {\em predicted} model shear for each configuration of galaxies.  Figure \ref{fig:three_pt_corr} illustrates the geometry on the sky.  Suppose that we have measurements of the shear at positions $\vec r_1$ and $\vec r_2$ relative to the center of the lens halo.  When we sum over all possible $\vec r_1$ and $\vec r_2$, the filter which weights each triangle is ${\bf F}(\vec r_1,\vec r_2) \propto \bm{\gamma}^{(2)}(\vec r_1)\otimes \bm{\gamma}^{(2)}(\vec r_2)$.  Because the orientation angle of the lens halo is unknown, the filter must depend on the relative position angle of the sources, not their absolute position angles: ${\bf F}(\vec r_1,\vec r_2)={\bf F}(|\vec r_1|,|\vec r_2|,\Delta\theta_{12})$.  We compute {\bf F} by averaging all possible triangles with one vertex at the center of the lens, and a constant opening angle, $\Delta\theta_{12}$, between position vectors to the lensed galaxies, of magnitude $\lvert \vec r_1\rvert$ and $\lvert \vec r_2\rvert$ (see Fig. \ref{fig:three_pt_corr}).  Since each shear has 2 components, the filter {\bf F} is a $2\times2$ matrix, with components
\begin{eqnarray} \label{eq:filter_element}
F_{ij}\left(r_1,r_2,\Delta\theta_{12}\right) &=& \left\langle g_i(\vec r_1)g_j(\vec r_2) \right\rangle \\
	&&\hspace{-2cm}= \frac{\int r_1' r_2' dr_1' dr_2' d\theta_1' d\theta_2' g_i(\vec r_1)g_j(\vec r_2)\delta_{r_1'}\delta_{r_2'}\delta_{\Delta\theta_{12}'}}{\int r_1' r_2' dr_1' dr_2' d\theta_1' d\theta_2' ~\delta_{r_1'}\delta_{r_2'}\delta_{\Delta\theta_{12}'}}  \nonumber,
\end{eqnarray}
where indices $i,j$ run over $+,\times$, and we define
$\Delta\theta_{12} =\theta_2-\theta_1 = \cos^{-1}(\vec r_1\cdot\vec r_2/r_1r_2)$, along with
$\delta_{r_\alpha'}\equiv\delta(r_\alpha'-r_\alpha)$ and $\delta_{\Delta\theta_{12}'}\equiv\delta(\Delta\theta_{12}'-\Delta\theta_{12})$.
The integral in \eqref{eq:filter_element} covers the projected area in the sky, where $r$ ranges from some $r_{\rm min}$ to $r_{\rm max}$,  and $\theta$ ranges from $0$ to $2\pi$. Simplifying using the $\delta$ functions, we obtain
\begin{eqnarray}
	F_{++} &=& \frac{1}{2} g_+(r_1) g_+(r_2)\cos 2\Delta\theta_{12} \nonumber \\
	F_{+\times} &=& \frac{1}{2} g_+(r_1) g_\times(r_2)\sin 2\Delta\theta_{12} \nonumber \\
	F_{\times +} &=& -\frac{1}{2}g_\times(r_1) g_+(r_2)\sin 2\Delta\theta_{12} \nonumber \\
	F_{\times\times} &=& \frac{1}{2} g_\times(r_1) g_\times(r_2)\cos 2\Delta\theta_{12}
\label{eq:F_elements}
\end{eqnarray}

Eqns.\ \eqref{eq:F_elements} specify the elements of the filter weighting each possible triangle in the 3-point correlation function.  We then evaluate our estimator by summing over all triangles, weighting the shear by {\bf F}.  Explicitly, we evaluate
\begin{equation} \label{eq:estimator_1}
f_{\rm obs}=\left\langle\boldsymbol{\gamma}(\vec r_1) \cdot \mathbf{F}(r_1,r_2,\Delta\theta_{12}) \cdot \boldsymbol{\gamma}(\vec r_2)\right\rangle,
\end{equation}
where the expectation value implies averaging over all possible $\vec r_1$ and $\vec r_2$.  Note that, because of the angular dependence of {\bf F}, Eqn.\ \eqref{eq:estimator_1} is only sensitive to the quadrupolar component of the shear.

In order to translate $f_{\rm obs}$ into an estimate for the ellipticity $\epsilon$, we need to know what result Eqn.\ \eqref{eq:estimator_1} will give as a function of $\epsilon$.  We can compute this by inserting the predicted model shear into the equation.  Specifically, let us define
\begin{align} \label{eq:estimator_analytic}
	& f_{\rm model}=\left\langle \bm{g}(\vec r_1) \cdot {\bf F} \cdot \bm{g}(\vec r_2)\right\rangle \\
	&=\frac{1}{2}\Big\langle g_+^2(r_1) g_+^2(r_2) \cos2\theta_1\cos2\theta_2\cos2\Delta\theta_{12} \notag \\ 
	&~~~~+ g_+^2(r_1) g_\times^2(r_2) \cos2\theta_1\sin2\theta_2\sin2\Delta\theta_{12} \notag \\
	&~~~~- g_\times^2(r_1) g_+^2(r_2) \sin2\theta_1\cos2\theta_2\sin2\Delta\theta_{12} \notag \\
	&~~~~+ g_\times^2(r_1) g_\times^2(r_2) \sin2\theta_1\sin2\theta_2\cos2\Delta\theta_{12}\Big\rangle	\notag \\
	&=\frac{1}{2}\left\{\frac{\pi}{A}\int_{r_{\rm min}}^{r_{\rm max}} \left[g_+^2(r) + g_\times^2(r)\right] r dr\right\}^2 \notag \\
	&= \frac{1}{2}\left\{\frac{\int_{r_{\rm min}}^{r_{\rm max}} \left[g_+^2(r) + g_\times^2(r)\right] r dr}{r_{\rm max}^2 - r_{\rm min}^2}\right\}^2 \notag .
\end{align}
 Then
\begin{equation}
\epsilon=\sqrt{\frac{f_{\rm obs}}{f_{\rm model}}}
\end{equation}
This defines our estimator for the ellipticity $\epsilon$.  To reiterate, the ingredient in our expression is the radial profile of the average monopole density profile $\langle\kappa_0\rangle(r)$, which may be reconstructed from the stacked tangential shear profile $\langle\gamma_+\rangle(r)$.  Given $\kappa_0(r)$, we may then determine the functions $g_+$ and $g_\times$ which enter the estimator.  In the next section, we apply this estimator to samples of halos from N-body simulations, to gauge how well we can measure halo ellipticities for realistic objects.

\section{Results}\label{sec:Results}

\begin{figure*}
\centering
\includegraphics[width=\textwidth, trim=0in 0in 0in .8in, clip]{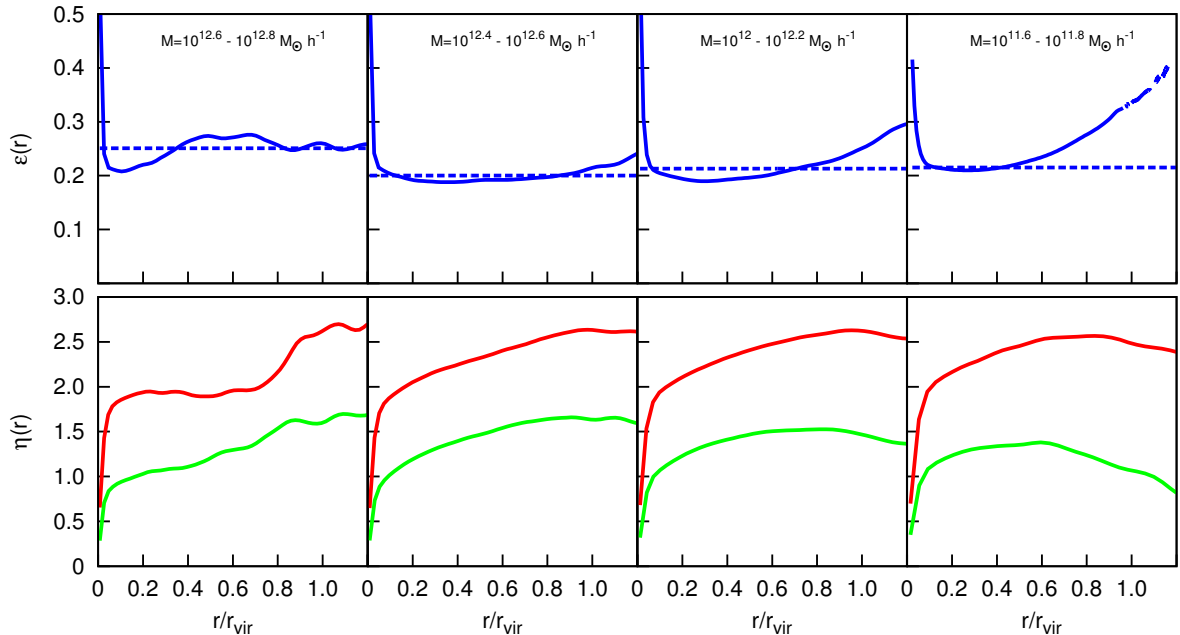}
\caption{The top panels show the comparison between the average ellipticities of halos in various mass bins (solid curve), compared to the ellipticity determined from the 3-point estimator (dashed line).  The bottom panel shows the local slope of the isotropic component (monopole) of the halos. The red curve corresponds to the smoothed slope of the 3D profile, and the green curve corresponds to the local slope of the projected 2D profile.  These slopes were measured from the stacked profiles after they were smoothed using a 6th order Savitzky-Golay filter over 17 nearest bins \citep{diemer2014dependence}. In the low mass bins, we observe significant departures from NFW slopes at large radii, possibly indicating the effects of nearby halos.}
\label{fig:nbody}
\end{figure*}

In \S\ref{sec:model}, we proposed a 3-point estimator for halo anisotropy.  In this section, we assess how well this estimator measures average halo ellipticities.  First, we use simulated halos from cosmological N-body simulations to quantify systematic errors caused by the fact that the structure of realistic halos will not be as simple as our monopole+quadrupole mass model.  Secondly, we quantify the statistical error associated with the finite number of lens-source-source triples.  Because each source galaxy provides an extremely noisy estimate of the shear, large numbers of triples are required to suppress this statistical shape noise. 

\subsection{Comparison with N-body simulations}

We have applied our estimator (Eqn. \ref{eq:estimator_1}) to halos from the publicly available Bolshoi simulation \citep{Klypin2011}.  Using the Rockstar catalog provided by the MultiDark database \footnote{http://hipacc.ucsc.edu/Bolshoi/MergerTrees.html}, we selected halos with virial masses in the range $10^{11.7}-10^{12.7} M_{\odot}h^{-1}$. We downloaded particles within $5r_{\rm vir}$ of the halo center, to account for the mass within the halo as well as the nearby neighborhood.  For each halo, we construct three projections, along the simulation box axes, to construct convergence and shear maps.  From these shear maps, we then apply our estimator to measure the halo ellipticity $\epsilon$.  Figure \ref{fig:nbody} shows the results of our measurement across several mass bins.  For comparison, the figure also plots the ellipticity directly measured from the projected mass profiles.  In all cases, we find good agreement, despite several potential systematics discussed below.

First, our mass model assumes that ellipticity $\epsilon$ is constant with radius, meaning that the shape of the quadrupole $\kappa_2(r)$ of the mass distribution may be determined from the shape of the monopole profile $\kappa_0(r)$.  For individual galaxies, the mass distributions are unknown.  Galaxy-galaxy lensing can be used to reconstruct the mean monopole profile $\langle\kappa_0\rangle(r)$, but individual halos will have radial profiles that vary from the mean.  Because our estimator is not linear in the shear, this scatter in radial profiles can bias our measurement.  To estimate the size of this potential bias, we generated artificial halos with radial profiles consistent with the Bolshoi halos (i.e.\ same $M_{\rm vir}$ and $c_{\rm vir}$) but with specified values of $\epsilon$.  For the range of concentrations found in the mass range we have considered, we find a {\em fractional} bias in the reconstructed $\epsilon$ of $\sim 3-6\%$.

\begin{figure}
    \centering
    \includegraphics[width=0.52\textwidth]{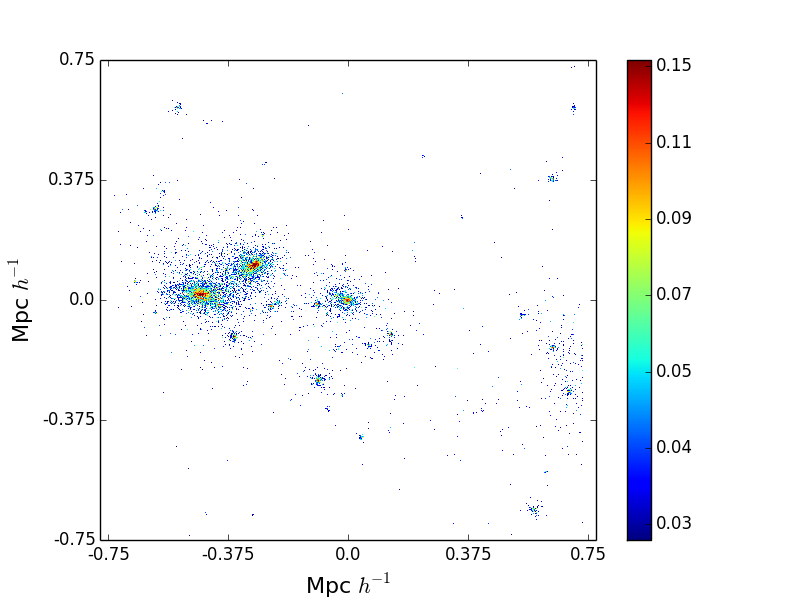}
    \caption{Example of line of sight projection effects.  The halo at the origin has massive neighbors projecting nearby, which generate a large quadrupole moment that is unrelated to the halo's own ellipticity. Colors correspond to convergence $\kappa$, for a lens redshift of $z_l=0.3$ and source redshift of $z_s=0.5$.  }
    \label{fig:proj}
\end{figure}

A second potential source of systematic errors arises from projections of other halos.  Our mass model assumes that all shear is generated by the halos hosting the stacked galaxies.  In reality, however, not all halos are isolated: other objects can project near the objects we are stacking and contaminate our measurement.  Figure \ref {fig:proj} shows one such example.  Such projections can produce very large quadrupoles (and other multipoles) near certain halos, and because our estimator is not linear in the shear, this contamination can bias our results.

In general, there are two types of projections relevant to our measurement: galaxies that are correlated with the foreground lenses, and uncorrelated galaxies that randomly project into the line of sight.  It is straightforward to correct for the uncorrelated projections.  We could, for example, simply stack on random sky points instead of lens galaxies, and subtract this from our estimator.   Mitigating the effects of correlated structures is not as easy.  Perhaps the simplest approach would be to stack only galaxies that are relatively isolated, i.e.\ galaxies that are clearly central galaxies (not satellites), and that have no comparably bright galaxies nearby the line of sight.  Fortunately, it appears that the bias due to projections of correlated structures may not be large.  In our calculations, we have not corrected for projections in any way, beyond stacking only on halos not subhalos (i.e.\ centrals, not satellites).  Our measurement is therefore contaminated by projections of other halos within 5 $r_{\rm vir}$ (as in Figure \ref{fig:proj}).  Arguably, this should account for most of the correlated objects.  The galaxy auto-correlation function behaves close to $\xi(r)\propto r^{-2}$ in 3D \citep{peebles}, so the number of galaxies with 3D radius $r>5r_{\rm vir}$ that project onto small radius should be about $\sim 1/5$ of the number of galaxies with 3D radius $r>r_{\rm vir}$ projecting onto small radius.  Because we have extracted particles out to $5r_{\rm vir}$, we should account for about $\sim80\%$ of the correlated projections.

Our calculations should therefore include most of the effect of projections of correlated structure, and as Figure \ref{fig:nbody} illustrates, the effect of those projections on the stacked profiles of central galaxies is likely to be small.  Only in the lowest mass bin ($M\approx 10^{11.7} M_\odot h^{-1}$) do we observe any effects of the 2-halo term, and even there the recovered $\epsilon$ from the 3-point estimator is consistent with the halo ellipticity measured over the radial range where the 1-halo term is dominant.  Nevertheless, when measuring halo ellipticity for real lenses, it will be important to restrict the analysis to the regime where the 1-halo term dominates, which may be determined by modeling the stacked tangential shear profile $\langle\gamma_+\rangle$.  Overall, our analysis of N-body halos suggests that systematic errors due to our simplistic mass model will not significantly bias our measurement of halo anisotropy.
        
\subsection{Shape noise}

In most regimes of weak lensing, the shear signal due to weak gravitational lensing is orders of magnitude weaker than the noise introduced by the intrinsic distribution of galaxy shapes and orientations.  To estimate the magnitude of the errors induced by shape noise, let us first define the shape noise per galaxy $\vec{\cal N} = \{{\cal N}_+,{\cal N}_\times\}$.  Each component of $\vec{\cal N}$ is assumed to be a Gaussian random variable with covariance
\begin{equation}
\langle {\cal N}_i {\cal N}_j \rangle=\sigma_{\epsilon}^2\delta_{ij}
\end{equation}
where the indices $i$ and $j$ correspond to the tangential and cross components of the noise, and $\sigma_{\epsilon}=0.25$ \citep{Chang}.  The number density of source galaxies is $n(\vec x)=\sum_i \delta(\vec x - \vec x_i)$, with mean number density $\bar n$.  Then the 2-point correlation of the shape noise is
\begin{equation}
\langle {\cal N}_i(\vec x_1) {\cal N}_j(\vec x_2) \rangle=\sigma_{\epsilon}^2\delta_{ij}\frac{\delta(\vec x_1-\vec x_2)}{\bar{n}}
\label{covar}
\end{equation}
Within an area $A$, where the number of galaxies is approximately $N=\bar{n}A$, the signal expectation value derived in Eqn.\ \eqref{eq:estimator_analytic} is
\begin{equation}
S = \langle\bm{\gamma}\cdot{\bf F}\cdot\bm{\gamma}\rangle = \frac{\pi^2\epsilon^2}{2A^2}\left\{\int{\left[g_+^2(r)+g_\times^2(r)\right]r\,dr}\right\}^2
\end{equation}
In comparison, the noise variance is
\begin{eqnarray}
\sigma_{N}^2&=&\langle (\mathbf{\mathcal{N}\cdot F \cdot {\cal N}})^2\rangle-\left(\langle{\cal N}\cdot{\bf F}\cdot{\cal N}\rangle\right)^2 \nonumber \\
 &=& \frac{\sigma_{\epsilon}^4 \pi^2}{\bar{n}^2 A^4}\left\{\int{\left[g_+^2(r)+g_\times^2(r)\right]r\,dr}\right\}^2
\end{eqnarray}
where the expectation value is computed by both summing over all possible triangles in the sky and also by taking the ensemble average of the Gaussian noise field.  In the second equality, we have used Eqn.\ \eqref{covar} and Wick's theorem.
Therefore the expected signal to noise per lens galaxy for the constructed estimator is
\begin{equation}\label{eq:snr}
\frac{S}{\sigma_{N}}= \frac{\pi\epsilon^2 \bar{n}}{2 \sigma_{\epsilon}^2}\int\left[g_+^2(r)+g_\times^2(r)\right]r\,dr.
\end{equation}
As we might expect, the signal to noise ratio (SNR) per lens scales quadratically in the shear.  Therefore the signal should be easiest to detect for more massive galaxies that produce stronger shear, as long as the abundance of galaxies does not fall steeply with mass.  To get a sense of the expected SNR, we can perform a rough estimate by approximating the halo profile as isothermal ($\Sigma_0\propto r^{-1}$), which Figure \ref{fig:nbody} shows is not a terrible approximation over the radial range of interest.  To be concrete, suppose that the monopole profile is $\kappa_0(r)=b/(2r)$.  Plugging this into Eqn.\ \eqref{eq:snr}, we find that $S/\sigma_N \approx [\pi b^2 \bar n \epsilon^2/(8\sigma_\epsilon^2)]\log(r_{\rm max}/r_{\rm min})$ per lens.  Taking $\epsilon=0.2$, $\sigma_\epsilon=0.25$, $\bar n = 12\, {\rm arcmin}^{-2}$ as appropriate for DES, $b=1$ arcsecond, and $r_{\rm max}/r_{\rm min}=20$ gives $S/\sigma_N \approx 0.0025$, meaning that with $10^6$ such lenses, we could detect the expected ellipticity at $\sim 2.5\sigma$.  LSST will have more than twice the effective number density of sources \citep{Chang}, more than doubling the signal to noise of the 3-point estimator.  At this point, it is perhaps worth comparing this estimate with the corresponding signal/noise ratio for a 2-point estimator.  Repeating the argument of \S\ref{sec:model} for the analogous 2-point estimator, we find that per lens, $(S/\sigma_N)_{\rm 2 pt} = \sqrt{2(S/\sigma_N)_{\rm 3 pt}}$, for halos that are perfectly aligned with their galaxies on the sky.  Since the SNR per lens is much less than 1, this illustrates that 2-point estimates of halo anisotropy will have much greater statistical sensitivity than 3-point estimators.  As noted above, however, this superior statistical power may be irrelevant if systematic effects due to halo misalignments remain uncertain.  

\section{Discussion}

We have shown that the lens-shear-shear three-point correlation function can be used to extract the ellipticity of dark matter halos, without the need to align the light profiles of galaxies that are being stacked.  Using a simple model of the projected surface density profiles of dark matter halos, we constructed an estimator for halo ellipticity that sums over all triangular configurations of the 3-point function.  We validated our estimator using simulated halos from the Bolshoi cosmological simulation, showing that the shear-derived estimator yields results consistent with the ellipticity measured directly from the particle data.  We investigated potential sources of systematic error, and argued that they should be small, at the $\sim 5\%$ level, well below theoretical uncertainties. We also estimated the signal to noise ratios expected for imaging surveys, and found that deep imaging surveys should be able to detect halo ellipticities.  The total signal to noise scales with the number of lens-source-source triplets as $N_t^{1/2}\propto n_l^{1/2} n_s$, meaning that deep imaging surveys with large effective number densities of sources will be most sensitive.  Ongoing surveys like PanSTARRS, DES, and HSC may be able to detect halo ellipticities, while future surveys like Euclid or LSST should have sufficient sensitivity for a significant detection.  The same surveys will, of course, be able to measure 2-point galaxy-galaxy lensing with far greater signal to noise than 3-point lensing.  However, if 2-point estimators are limited by systematic uncertainties, as suggested by theoretical work on galaxy-halo misalignments \citep{Bett2012}, then 3-point lensing could prove to be a useful probe of halo anisotropy.

In \S\ref{sec:Results}, we discussed potential theoretical systematic errors, which could arise if our simple mass model failed to describe actual halos adequately.  Besides theoretical systematics, our proposed measurement will also be liable to possible observational systematics.  One obvious observational source of systematic error is point spread function (PSF) anisotropy.  The PSF determines how the actual shape of a galaxy on the sky is related to the observed shape of a galaxy, measured by a camera on a telescope possibly beneath the distorting effects of the Earth's atmosphere.  Our ability to measure the PSF is frequently a limiting factor in our ability to measure the true shapes of weakly lensed galaxies, which degrades our ability to measure shear.  In principle, this could be disastrous for the halo ellipticity measurement we have proposed.  For example, if the PSF were uniformly anisotropic across the virial radius of a lens halo, leading to a spurious, uniform shear, this would exactly mimic the ellipticity signal we are seeking to detect.  In practice, however, mitigating such effects in galaxy-galaxy lensing measurements should be straightforward, as long as the shape of the PSF is not strongly correlated with the number of foreground lens galaxies.  For example, we can assess the extent of such PSF anisotropies by stacking on random sky points instead of lens galaxies.  Even if PSF anisotropies are present, the ellipticity signal should appear as an excess correlation with lens galaxies, above what is seen around random sky points.  As discussed in \S\ref{sec:Results}, the same test would also help remove the effects of masses uncorrelated with the lens galaxies.

Another potential astrophysical contamination of the signal arises from intrinsic alignment between galaxies \citep{2000ApJ...545..561C,2000MNRAS.319..649H,2001ApJ...559..552C,2004PhRvD..70f3526H,2004MNRAS.353..529H,2006MNRAS.367..611M, 2007MNRAS.381.1197H}. Galaxies that form and evolve in the same local environment may be systematically aligned with each other due to long range tidal effects \citep{Blazek2011}, consequently replicating the correlation that is produced by gravitational lensing.  This effect can manifest itself in two ways, (i) nearby source galaxies can be preferentially aligned with each other, and (ii) lens-source pairs can be physically associated with each other, if (for example) a fraction of source galaxies are satellites of the lensing, foreground galaxy.  Both these problems can be mitigated using redshift information, for example by excluding galaxy pairs with similar redshifts.  Observationally, the contamination of galaxy-galaxy lensing due to alignments from lens-source correlations produced by photometric redshift errors has been shown to be exceedingly small in SDSS \citep{Blazek2012}.   Stacking on random points, as discussed above, would also help to quantify and remove the effect of alignments of source galaxies.

Magnification of lenses could also produce a systematic effect on the signal.  Our estimator correlates the number density of foreground galaxies, $n_g$, to the shear at two positions in the sky, $\gamma_1$ and $\gamma_2$. The foreground galaxies (lenses) are lensed by matter distribution between the observer and the lens along the line of sight. This causes a modification of the clustering of lenses due to cosmic magification along the line of sight.  The variation from the unlensed number density is, to lowest order, linear in the lensing convergence, $\kappa_{<}$ \cite{Simon2013}. In addition the shear itself has a contribution from the matter density, integrated along the line of sight to the redshift of the source. The combined effect therefore contributes to the 3-point correlator, $\langle n_{g}'\gamma_1'\gamma_2'\rangle$, terms like $\langle\kappa_{<}~\gamma_{1<}~\gamma_{2<}\rangle$. These third order shear correlations have been measured to be less than $10^{-7}$ \citep{Jarvis2004, Semboloni2011} for aperture scales of $\theta \sim 1'$ and source redshift $z_s\sim 1$, while \citet{Simon2013} predict an upper bound to the effect of magnification of lenses on three point statistics of $10^{-8}$ for sources at $z_s\sim 0.4$. Therefore, it appears that this effect will not significantly contaminate the measurement of halo ellipticity.

Another potential systematic is the effect of baryons, which can act to modify the halo axis ratios on small scales $\lesssim 0.25 r_{\rm vir}$ \citep{Kazantzidis2004,Bryan2013}.  Judging from Figure \ref{fig:nbody}, our estimator is most sensitive to the ellipticity at somewhat larger radii, suggesting that baryonic effects will be limited.  In principle, we can suppress any baryonic effects on our estimator by restricting the range of integration ($r_{\rm min}$ and $r_{\rm max}$ in Eqns.\ \eqref{eq:filter_element}-\eqref{eq:estimator_1}) to exclude regions that may be contaminated.  
Alternatively, given sufficient signal to noise, one could try to measure the ellipticity as a function of radius by subdividing the sample, for example by comparing triangles at large vs.\ small separation. Besides constraining any radial variation in ellipticity, 3-point lensing could also probe any misalignments between the principal axes at small radii vs.\ large radii.  We defer such possibilities to future work.

In this paper, we have investigated one particular application of the measurement of 3-point correlation functions.  As theoretical work has shown \citep{Simon2013}, high-order correlation functions contain significant amounts of information above and beyond that encoded in better studied 2-point functions.  The advent of deep, wide-area imaging surveys is now making the measurement of these high-order correlations practical across a range of spatial scales, suggesting that this will be a fruitful area of research for years to come.

\begin{acknowledgments}
ND is supported by NASA under grants NNX12AD02G and NNX12AC99G, and by a Sloan Fellowship.
The MultiDark Database used in this paper and the web application providing online access to it
were constructed as part of the activities of the German Astrophysical Virtual Observatory as result
of a collaboration between the Leibniz-Institute for Astrophysics Potsdam (AIP) and the Spanish
MultiDark Consolider Project CSD2009-00064. The Bolshoi and MultiDark simulations were run on the
NASA's Pleiades supercomputer at the NASA Ames Research Center.
\end{acknowledgments}

\newcommand{\apjl}{\apj\ Lett.}
\newcommand{\jcap}{Journal of Cosmology and Astroparticle Physics}
\newcommand{\mnras}{Monthly Notices of the Royal Astronomical Society}

\bibliography{paper}

\begin{thebibliography}{48}%
\makeatletter
\providecommand \@ifxundefined [1]{%
 \@ifx{#1\undefined}
}%
\providecommand \@ifnum [1]{%
 \ifnum #1\expandafter \@firstoftwo
 \else \expandafter \@secondoftwo
 \fi
}%
\providecommand \@ifx [1]{%
 \ifx #1\expandafter \@firstoftwo
 \else \expandafter \@secondoftwo
 \fi
}%
\providecommand \natexlab [1]{#1}%
\providecommand \enquote  [1]{``#1''}%
\providecommand \bibnamefont  [1]{#1}%
\providecommand \bibfnamefont [1]{#1}%
\providecommand \citenamefont [1]{#1}%
\providecommand \href@noop [0]{\@secondoftwo}%
\providecommand \href [0]{\begingroup \@sanitize@url \@href}%
\providecommand \@href[1]{\@@startlink{#1}\@@href}%
\providecommand \@@href[1]{\endgroup#1\@@endlink}%
\providecommand \@sanitize@url [0]{\catcode `\\12\catcode `\$12\catcode
  `\&12\catcode `\#12\catcode `\^12\catcode `\_12\catcode `\%12\relax}%
\providecommand \@@startlink[1]{}%
\providecommand \@@endlink[0]{}%
\providecommand \url  [0]{\begingroup\@sanitize@url \@url }%
\providecommand \@url [1]{\endgroup\@href {#1}{\urlprefix }}%
\providecommand \urlprefix  [0]{URL }%
\providecommand \Eprint [0]{\href }%
\providecommand \doibase [0]{http://dx.doi.org/}%
\providecommand \selectlanguage [0]{\@gobble}%
\providecommand \bibinfo  [0]{\@secondoftwo}%
\providecommand \bibfield  [0]{\@secondoftwo}%
\providecommand \translation [1]{[#1]}%
\providecommand \BibitemOpen [0]{}%
\providecommand \bibitemStop [0]{}%
\providecommand \bibitemNoStop [0]{.\EOS\space}%
\providecommand \EOS [0]{\spacefactor3000\relax}%
\providecommand \BibitemShut  [1]{\csname bibitem#1\endcsname}%
\let\auto@bib@innerbib\@empty
\bibitem [{\citenamefont {{Dubinski}}\ and\ \citenamefont
  {{Carlberg}}(1991)}]{Dubinski1991}%
  \BibitemOpen
  \bibfield  {author} {\bibinfo {author} {\bibfnamefont {J.}~\bibnamefont
  {{Dubinski}}}\ and\ \bibinfo {author} {\bibfnamefont {R.~G.}\ \bibnamefont
  {{Carlberg}}},\ }\href {\doibase 10.1086/170451} {\bibfield  {journal}
  {\bibinfo  {journal} {\apj}\ }\textbf {\bibinfo {volume} {378}},\ \bibinfo
  {pages} {496} (\bibinfo {year} {1991})}\BibitemShut {NoStop}%
\bibitem [{\citenamefont {{Jing}}\ and\ \citenamefont
  {{Suto}}(2002)}]{Jing2002}%
  \BibitemOpen
  \bibfield  {author} {\bibinfo {author} {\bibfnamefont {Y.~P.}\ \bibnamefont
  {{Jing}}}\ and\ \bibinfo {author} {\bibfnamefont {Y.}~\bibnamefont
  {{Suto}}},\ }\href {\doibase 10.1086/341065} {\bibfield  {journal} {\bibinfo
  {journal} {\apj}\ }\textbf {\bibinfo {volume} {574}},\ \bibinfo {pages} {538}
  (\bibinfo {year} {2002})},\ \Eprint {http://arxiv.org/abs/astro-ph/0202064}
  {astro-ph/0202064} \BibitemShut {NoStop}%
\bibitem [{\citenamefont {{Allgood}}\ \emph {et~al.}(2006)\citenamefont
  {{Allgood}}, \citenamefont {{Flores}}, \citenamefont {{Primack}},
  \citenamefont {{Kravtsov}}, \citenamefont {{Wechsler}}, \citenamefont
  {{Faltenbacher}},\ and\ \citenamefont {{Bullock}}}]{Allgood2006}%
  \BibitemOpen
  \bibfield  {author} {\bibinfo {author} {\bibfnamefont {B.}~\bibnamefont
  {{Allgood}}}, \bibinfo {author} {\bibfnamefont {R.~A.}\ \bibnamefont
  {{Flores}}}, \bibinfo {author} {\bibfnamefont {J.~R.}\ \bibnamefont
  {{Primack}}}, \bibinfo {author} {\bibfnamefont {A.~V.}\ \bibnamefont
  {{Kravtsov}}}, \bibinfo {author} {\bibfnamefont {R.~H.}\ \bibnamefont
  {{Wechsler}}}, \bibinfo {author} {\bibfnamefont {A.}~\bibnamefont
  {{Faltenbacher}}}, \ and\ \bibinfo {author} {\bibfnamefont {J.~S.}\
  \bibnamefont {{Bullock}}},\ }\href {\doibase
  10.1111/j.1365-2966.2006.10094.x} {\bibfield  {journal} {\bibinfo  {journal}
  {\mnras}\ }\textbf {\bibinfo {volume} {367}},\ \bibinfo {pages} {1781}
  (\bibinfo {year} {2006})},\ \Eprint {http://arxiv.org/abs/astro-ph/0508497}
  {astro-ph/0508497} \BibitemShut {NoStop}%
\bibitem [{\citenamefont {{Schneider}}\ \emph {et~al.}(2012)\citenamefont
  {{Schneider}}, \citenamefont {{Frenk}},\ and\ \citenamefont
  {{Cole}}}]{Schneider2012}%
  \BibitemOpen
  \bibfield  {author} {\bibinfo {author} {\bibfnamefont {M.~D.}\ \bibnamefont
  {{Schneider}}}, \bibinfo {author} {\bibfnamefont {C.~S.}\ \bibnamefont
  {{Frenk}}}, \ and\ \bibinfo {author} {\bibfnamefont {S.}~\bibnamefont
  {{Cole}}},\ }\href {\doibase 10.1088/1475-7516/2012/05/030} {\bibfield
  {journal} {\bibinfo  {journal} {\jcap}\ }\textbf {\bibinfo {volume} {5}},\
  \bibinfo {eid} {030} (\bibinfo {year} {2012})},\ \Eprint
  {http://arxiv.org/abs/1111.5616} {arXiv:1111.5616 [astro-ph.CO]} \BibitemShut
  {NoStop}%
\bibitem [{\citenamefont {{Dav\'e}}\ \emph {et~al.}(2001)\citenamefont
  {{Dav\'e}}, \citenamefont {{Spergel}}, \citenamefont {{Steinhardt}},\ and\
  \citenamefont {{Wandelt}}}]{Dave2001}%
  \BibitemOpen
  \bibfield  {author} {\bibinfo {author} {\bibfnamefont {R.}~\bibnamefont
  {{Dav\'e}}}, \bibinfo {author} {\bibfnamefont {D.~N.}\ \bibnamefont
  {{Spergel}}}, \bibinfo {author} {\bibfnamefont {P.~J.}\ \bibnamefont
  {{Steinhardt}}}, \ and\ \bibinfo {author} {\bibfnamefont {B.~D.}\
  \bibnamefont {{Wandelt}}},\ }\href@noop {} {\bibfield  {journal} {\bibinfo
  {journal} {\apj}\ }\textbf {\bibinfo {volume} {547}},\ \bibinfo {pages} {574}
  (\bibinfo {year} {2001})}\BibitemShut {NoStop}%
\bibitem [{\citenamefont {{Peter}}\ \emph {et~al.}(2013)\citenamefont
  {{Peter}}, \citenamefont {{Rocha}}, \citenamefont {{Bullock}},\ and\
  \citenamefont {{Kaplinghat}}}]{Peter13}%
  \BibitemOpen
  \bibfield  {author} {\bibinfo {author} {\bibfnamefont {A.~H.~G.}\
  \bibnamefont {{Peter}}}, \bibinfo {author} {\bibfnamefont {M.}~\bibnamefont
  {{Rocha}}}, \bibinfo {author} {\bibfnamefont {J.~S.}\ \bibnamefont
  {{Bullock}}}, \ and\ \bibinfo {author} {\bibfnamefont {M.}~\bibnamefont
  {{Kaplinghat}}},\ }\href {\doibase 10.1093/mnras/sts535} {\bibfield
  {journal} {\bibinfo  {journal} {\mnras}\ }\textbf {\bibinfo {volume} {430}},\
  \bibinfo {pages} {105} (\bibinfo {year} {2013})},\ \Eprint
  {http://arxiv.org/abs/1208.3026} {arXiv:1208.3026 [astro-ph.CO]} \BibitemShut
  {NoStop}%
\bibitem [{\citenamefont {{Rocha}}\ \emph {et~al.}(2013)\citenamefont
  {{Rocha}}, \citenamefont {{Peter}}, \citenamefont {{Bullock}}, \citenamefont
  {{Kaplinghat}}, \citenamefont {{Garrison-Kimmel}}, \citenamefont
  {{O{\~n}orbe}},\ and\ \citenamefont {{Moustakas}}}]{Rocha13}%
  \BibitemOpen
  \bibfield  {author} {\bibinfo {author} {\bibfnamefont {M.}~\bibnamefont
  {{Rocha}}}, \bibinfo {author} {\bibfnamefont {A.~H.~G.}\ \bibnamefont
  {{Peter}}}, \bibinfo {author} {\bibfnamefont {J.~S.}\ \bibnamefont
  {{Bullock}}}, \bibinfo {author} {\bibfnamefont {M.}~\bibnamefont
  {{Kaplinghat}}}, \bibinfo {author} {\bibfnamefont {S.}~\bibnamefont
  {{Garrison-Kimmel}}}, \bibinfo {author} {\bibfnamefont {J.}~\bibnamefont
  {{O{\~n}orbe}}}, \ and\ \bibinfo {author} {\bibfnamefont {L.~A.}\
  \bibnamefont {{Moustakas}}},\ }\href {\doibase 10.1093/mnras/sts514}
  {\bibfield  {journal} {\bibinfo  {journal} {\mnras}\ }\textbf {\bibinfo
  {volume} {430}},\ \bibinfo {pages} {81} (\bibinfo {year} {2013})},\ \Eprint
  {http://arxiv.org/abs/1208.3025} {arXiv:1208.3025 [astro-ph.CO]} \BibitemShut
  {NoStop}%
\bibitem [{\citenamefont {{Kaplinghat}}\ \emph {et~al.}(2013)\citenamefont
  {{Kaplinghat}}, \citenamefont {{Keeley}}, \citenamefont {{Linden}},\ and\
  \citenamefont {{Yu}}}]{2013arXiv1311.6524K}%
  \BibitemOpen
  \bibfield  {author} {\bibinfo {author} {\bibfnamefont {M.}~\bibnamefont
  {{Kaplinghat}}}, \bibinfo {author} {\bibfnamefont {R.~E.}\ \bibnamefont
  {{Keeley}}}, \bibinfo {author} {\bibfnamefont {T.}~\bibnamefont {{Linden}}},
  \ and\ \bibinfo {author} {\bibfnamefont {H.-B.}\ \bibnamefont {{Yu}}},\
  }\href@noop {} {\bibfield  {journal} {\bibinfo  {journal} {ArXiv e-prints}\ }
  (\bibinfo {year} {2013})},\ \Eprint {http://arxiv.org/abs/1311.6524}
  {arXiv:1311.6524 [astro-ph.CO]} \BibitemShut {NoStop}%
\bibitem [{\citenamefont {{Helmi}}(2004)}]{Helmi2004}%
  \BibitemOpen
  \bibfield  {author} {\bibinfo {author} {\bibfnamefont {A.}~\bibnamefont
  {{Helmi}}},\ }\href@noop {} {\bibfield  {journal} {\bibinfo  {journal}
  {\apj}\ }\textbf {\bibinfo {volume} {610}},\ \bibinfo {pages} {L97} (\bibinfo
  {year} {2004})}\BibitemShut {NoStop}%
\bibitem [{\citenamefont {{Law}}\ and\ \citenamefont
  {{Majewski}}(2010)}]{Law2010}%
  \BibitemOpen
  \bibfield  {author} {\bibinfo {author} {\bibfnamefont {D.~R.}\ \bibnamefont
  {{Law}}}\ and\ \bibinfo {author} {\bibfnamefont {S.~R.}\ \bibnamefont
  {{Majewski}}},\ }\href@noop {} {\bibfield  {journal} {\bibinfo  {journal}
  {\apj}\ }\textbf {\bibinfo {volume} {714}},\ \bibinfo {pages} {229} (\bibinfo
  {year} {2010})}\BibitemShut {NoStop}%
\bibitem [{\citenamefont {{Deg}}\ and\ \citenamefont
  {{Widrow}}(2013)}]{Deg2013}%
  \BibitemOpen
  \bibfield  {author} {\bibinfo {author} {\bibfnamefont {N.}~\bibnamefont
  {{Deg}}}\ and\ \bibinfo {author} {\bibfnamefont {L.}~\bibnamefont
  {{Widrow}}},\ }\href@noop {} {\bibfield  {journal} {\bibinfo  {journal}
  {Monthly Notices of the Royal Astronomical Society}\ }\textbf {\bibinfo
  {volume} {428}},\ \bibinfo {pages} {912} (\bibinfo {year}
  {2013})}\BibitemShut {NoStop}%
\bibitem [{\citenamefont {{Ibata}}\ \emph {et~al.}(2013)\citenamefont
  {{Ibata}}, \citenamefont {{Lewis}}, \citenamefont {{Martin}}, \citenamefont
  {{Bellazzini}},\ and\ \citenamefont {{Correnti}}}]{Ibata2013}%
  \BibitemOpen
  \bibfield  {author} {\bibinfo {author} {\bibfnamefont {R.}~\bibnamefont
  {{Ibata}}}, \bibinfo {author} {\bibfnamefont {G.~F.}\ \bibnamefont
  {{Lewis}}}, \bibinfo {author} {\bibfnamefont {N.~F.}\ \bibnamefont
  {{Martin}}}, \bibinfo {author} {\bibfnamefont {M.}~\bibnamefont
  {{Bellazzini}}}, \ and\ \bibinfo {author} {\bibfnamefont {M.}~\bibnamefont
  {{Correnti}}},\ }\href@noop {} {\bibfield  {journal} {\bibinfo  {journal}
  {\apj\ Lett.}\ }\textbf {\bibinfo {volume} {765}},\ \bibinfo {pages} {L15}
  (\bibinfo {year} {2013})}\BibitemShut {NoStop}%
\bibitem [{\citenamefont {{Vera-Ciro}}\ and\ \citenamefont
  {{Helmi}}(2013)}]{VeraCiro2013}%
  \BibitemOpen
  \bibfield  {author} {\bibinfo {author} {\bibfnamefont {C.}~\bibnamefont
  {{Vera-Ciro}}}\ and\ \bibinfo {author} {\bibfnamefont {A.}~\bibnamefont
  {{Helmi}}},\ }\href@noop {} {\bibfield  {journal} {\bibinfo  {journal} {\apj\
  Lett.}\ }\textbf {\bibinfo {volume} {773}},\ \bibinfo {pages} {L4} (\bibinfo
  {year} {2013})}\BibitemShut {NoStop}%
\bibitem [{\citenamefont {{van de Ven}}\ \emph {et~al.}(2010)\citenamefont
  {{van de Ven}}, \citenamefont {{Falc{\'o}n-Barroso}}, \citenamefont
  {{McDermid}}, \citenamefont {{Cappellari}}, \citenamefont {{Miller}},\ and\
  \citenamefont {{de Zeeuw}}}]{vandeVen2010}%
  \BibitemOpen
  \bibfield  {author} {\bibinfo {author} {\bibfnamefont {G.}~\bibnamefont {{van
  de Ven}}}, \bibinfo {author} {\bibfnamefont {J.}~\bibnamefont
  {{Falc{\'o}n-Barroso}}}, \bibinfo {author} {\bibfnamefont {R.~M.}\
  \bibnamefont {{McDermid}}}, \bibinfo {author} {\bibfnamefont
  {M.}~\bibnamefont {{Cappellari}}}, \bibinfo {author} {\bibfnamefont {B.~W.}\
  \bibnamefont {{Miller}}}, \ and\ \bibinfo {author} {\bibfnamefont {P.~T.}\
  \bibnamefont {{de Zeeuw}}},\ }\href {\doibase 10.1088/0004-637X/719/2/1481}
  {\bibfield  {journal} {\bibinfo  {journal} {\apj}\ }\textbf {\bibinfo
  {volume} {719}},\ \bibinfo {pages} {1481} (\bibinfo {year} {2010})},\ \Eprint
  {http://arxiv.org/abs/0807.4175} {arXiv:0807.4175} \BibitemShut {NoStop}%
\bibitem [{\citenamefont {{Suyu}}\ \emph {et~al.}(2012)\citenamefont {{Suyu}},
  \citenamefont {{Hensel}}, \citenamefont {{McKean}}, \citenamefont
  {{Fassnacht}}, \citenamefont {{Treu}}, \citenamefont {{Halkola}},
  \citenamefont {{Norbury}}, \citenamefont {{Jackson}}, \citenamefont
  {{Schneider}}, \citenamefont {{Thompson}}, \citenamefont {{Auger}},
  \citenamefont {{Koopmans}},\ and\ \citenamefont {{Matthews}}}]{Suyu2012}%
  \BibitemOpen
  \bibfield  {author} {\bibinfo {author} {\bibfnamefont {S.~H.}\ \bibnamefont
  {{Suyu}}}, \bibinfo {author} {\bibfnamefont {S.~W.}\ \bibnamefont
  {{Hensel}}}, \bibinfo {author} {\bibfnamefont {J.~P.}\ \bibnamefont
  {{McKean}}}, \bibinfo {author} {\bibfnamefont {C.~D.}\ \bibnamefont
  {{Fassnacht}}}, \bibinfo {author} {\bibfnamefont {T.}~\bibnamefont {{Treu}}},
  \bibinfo {author} {\bibfnamefont {A.}~\bibnamefont {{Halkola}}}, \bibinfo
  {author} {\bibfnamefont {M.}~\bibnamefont {{Norbury}}}, \bibinfo {author}
  {\bibfnamefont {N.}~\bibnamefont {{Jackson}}}, \bibinfo {author}
  {\bibfnamefont {P.}~\bibnamefont {{Schneider}}}, \bibinfo {author}
  {\bibfnamefont {D.}~\bibnamefont {{Thompson}}}, \bibinfo {author}
  {\bibfnamefont {M.~W.}\ \bibnamefont {{Auger}}}, \bibinfo {author}
  {\bibfnamefont {L.~V.~E.}\ \bibnamefont {{Koopmans}}}, \ and\ \bibinfo
  {author} {\bibfnamefont {K.}~\bibnamefont {{Matthews}}},\ }\href {\doibase
  10.1088/0004-637X/750/1/10} {\bibfield  {journal} {\bibinfo  {journal}
  {\apj}\ }\textbf {\bibinfo {volume} {750}},\ \bibinfo {eid} {10} (\bibinfo
  {year} {2012})},\ \Eprint {http://arxiv.org/abs/1110.2536} {arXiv:1110.2536
  [astro-ph.CO]} \BibitemShut {NoStop}%
\bibitem [{\citenamefont {{Bailin}}\ \emph {et~al.}(2008)\citenamefont
  {{Bailin}}, \citenamefont {{Power}}, \citenamefont {{Norberg}}, \citenamefont
  {{Zaritsky}},\ and\ \citenamefont {{Gibson}}}]{Bailin2008}%
  \BibitemOpen
  \bibfield  {author} {\bibinfo {author} {\bibfnamefont {J.}~\bibnamefont
  {{Bailin}}}, \bibinfo {author} {\bibfnamefont {C.}~\bibnamefont {{Power}}},
  \bibinfo {author} {\bibfnamefont {P.}~\bibnamefont {{Norberg}}}, \bibinfo
  {author} {\bibfnamefont {D.}~\bibnamefont {{Zaritsky}}}, \ and\ \bibinfo
  {author} {\bibfnamefont {B.~K.}\ \bibnamefont {{Gibson}}},\ }\href {\doibase
  10.1111/j.1365-2966.2008.13828.x} {\bibfield  {journal} {\bibinfo  {journal}
  {\mnras}\ }\textbf {\bibinfo {volume} {390}},\ \bibinfo {pages} {1133}
  (\bibinfo {year} {2008})},\ \Eprint {http://arxiv.org/abs/0706.1350}
  {arXiv:0706.1350} \BibitemShut {NoStop}%
\bibitem [{\citenamefont {{Sheldon}}\ \emph {et~al.}(2004)\citenamefont
  {{Sheldon}}, \citenamefont {{Johnston}}, \citenamefont {{Frieman}},
  \citenamefont {{Scranton}}, \citenamefont {{McKay}}, \citenamefont
  {{Connolly}}, \citenamefont {{Budav{\'a}ri}}, \citenamefont {{Zehavi}},
  \citenamefont {{Bahcall}}, \citenamefont {{Brinkmann}},\ and\ \citenamefont
  {{Fukugita}}}]{Sheldon2004}%
  \BibitemOpen
  \bibfield  {author} {\bibinfo {author} {\bibfnamefont {E.~S.}\ \bibnamefont
  {{Sheldon}}}, \bibinfo {author} {\bibfnamefont {D.~E.}\ \bibnamefont
  {{Johnston}}}, \bibinfo {author} {\bibfnamefont {J.~A.}\ \bibnamefont
  {{Frieman}}}, \bibinfo {author} {\bibfnamefont {R.}~\bibnamefont
  {{Scranton}}}, \bibinfo {author} {\bibfnamefont {T.~A.}\ \bibnamefont
  {{McKay}}}, \bibinfo {author} {\bibfnamefont {A.~J.}\ \bibnamefont
  {{Connolly}}}, \bibinfo {author} {\bibfnamefont {T.}~\bibnamefont
  {{Budav{\'a}ri}}}, \bibinfo {author} {\bibfnamefont {I.}~\bibnamefont
  {{Zehavi}}}, \bibinfo {author} {\bibfnamefont {N.~A.}\ \bibnamefont
  {{Bahcall}}}, \bibinfo {author} {\bibfnamefont {J.}~\bibnamefont
  {{Brinkmann}}}, \ and\ \bibinfo {author} {\bibfnamefont {M.}~\bibnamefont
  {{Fukugita}}},\ }\href {\doibase 10.1086/383293} {\bibfield  {journal}
  {\bibinfo  {journal} {The Astronomical Journal}\ }\textbf {\bibinfo {volume}
  {127}},\ \bibinfo {pages} {2544} (\bibinfo {year} {2004})},\ \Eprint
  {http://arxiv.org/abs/astro-ph/0312036} {astro-ph/0312036} \BibitemShut
  {NoStop}%
\bibitem [{\citenamefont {{Mandelbaum}}\ \emph
  {et~al.}(2006{\natexlab{a}})\citenamefont {{Mandelbaum}}, \citenamefont
  {{Seljak}}, \citenamefont {{Cool}}, \citenamefont {{Blanton}}, \citenamefont
  {{Hirata}},\ and\ \citenamefont {{Brinkmann}}}]{Mandelbaum06}%
  \BibitemOpen
  \bibfield  {author} {\bibinfo {author} {\bibfnamefont {R.}~\bibnamefont
  {{Mandelbaum}}}, \bibinfo {author} {\bibfnamefont {U.}~\bibnamefont
  {{Seljak}}}, \bibinfo {author} {\bibfnamefont {R.~J.}\ \bibnamefont
  {{Cool}}}, \bibinfo {author} {\bibfnamefont {M.}~\bibnamefont {{Blanton}}},
  \bibinfo {author} {\bibfnamefont {C.~M.}\ \bibnamefont {{Hirata}}}, \ and\
  \bibinfo {author} {\bibfnamefont {J.}~\bibnamefont {{Brinkmann}}},\ }\href
  {\doibase 10.1111/j.1365-2966.2006.10906.x} {\bibfield  {journal} {\bibinfo
  {journal} {\mnras}\ }\textbf {\bibinfo {volume} {372}},\ \bibinfo {pages}
  {758} (\bibinfo {year} {2006}{\natexlab{a}})},\ \Eprint
  {http://arxiv.org/abs/astro-ph/0605476} {astro-ph/0605476} \BibitemShut
  {NoStop}%
\bibitem [{\citenamefont {{Johnston}}\ \emph {et~al.}(2007)\citenamefont
  {{Johnston}}, \citenamefont {{Sheldon}}, \citenamefont {{Wechsler}},
  \citenamefont {{Rozo}}, \citenamefont {{Koester}}, \citenamefont {{Frieman}},
  \citenamefont {{McKay}}, \citenamefont {{Evrard}}, \citenamefont {{Becker}},\
  and\ \citenamefont {{Annis}}}]{Johnston2007}%
  \BibitemOpen
  \bibfield  {author} {\bibinfo {author} {\bibfnamefont {D.~E.}\ \bibnamefont
  {{Johnston}}}, \bibinfo {author} {\bibfnamefont {E.~S.}\ \bibnamefont
  {{Sheldon}}}, \bibinfo {author} {\bibfnamefont {R.~H.}\ \bibnamefont
  {{Wechsler}}}, \bibinfo {author} {\bibfnamefont {E.}~\bibnamefont {{Rozo}}},
  \bibinfo {author} {\bibfnamefont {B.~P.}\ \bibnamefont {{Koester}}}, \bibinfo
  {author} {\bibfnamefont {J.~A.}\ \bibnamefont {{Frieman}}}, \bibinfo {author}
  {\bibfnamefont {T.~A.}\ \bibnamefont {{McKay}}}, \bibinfo {author}
  {\bibfnamefont {A.~E.}\ \bibnamefont {{Evrard}}}, \bibinfo {author}
  {\bibfnamefont {M.~R.}\ \bibnamefont {{Becker}}}, \ and\ \bibinfo {author}
  {\bibfnamefont {J.}~\bibnamefont {{Annis}}},\ }\href@noop {} {\bibfield
  {journal} {\bibinfo  {journal} {ArXiv e-prints}\ } (\bibinfo {year}
  {2007})},\ \Eprint {http://arxiv.org/abs/0709.1159} {arXiv:0709.1159}
  \BibitemShut {NoStop}%
\bibitem [{\citenamefont {{Hoekstra}}\ \emph {et~al.}(2004)\citenamefont
  {{Hoekstra}}, \citenamefont {{Yee}},\ and\ \citenamefont
  {{Gladders}}}]{Hoekstra2004}%
  \BibitemOpen
  \bibfield  {author} {\bibinfo {author} {\bibfnamefont {H.}~\bibnamefont
  {{Hoekstra}}}, \bibinfo {author} {\bibfnamefont {H.~K.~C.}\ \bibnamefont
  {{Yee}}}, \ and\ \bibinfo {author} {\bibfnamefont {M.~D.}\ \bibnamefont
  {{Gladders}}},\ }\href@noop {} {\bibfield  {journal} {\bibinfo  {journal}
  {\apj}\ }\textbf {\bibinfo {volume} {606}},\ \bibinfo {pages} {67} (\bibinfo
  {year} {2004})}\BibitemShut {NoStop}%
\bibitem [{\citenamefont {{Mandelbaum}}\ \emph
  {et~al.}(2006{\natexlab{b}})\citenamefont {{Mandelbaum}}, \citenamefont
  {{Hirata}}, \citenamefont {{Broderick}}, \citenamefont {{Seljak}},\ and\
  \citenamefont {{Brinkmann}}}]{Mandelbaum2006}%
  \BibitemOpen
  \bibfield  {author} {\bibinfo {author} {\bibfnamefont {R.}~\bibnamefont
  {{Mandelbaum}}}, \bibinfo {author} {\bibfnamefont {C.~M.}\ \bibnamefont
  {{Hirata}}}, \bibinfo {author} {\bibfnamefont {T.}~\bibnamefont
  {{Broderick}}}, \bibinfo {author} {\bibfnamefont {U.}~\bibnamefont
  {{Seljak}}}, \ and\ \bibinfo {author} {\bibfnamefont {J.}~\bibnamefont
  {{Brinkmann}}},\ }\href@noop {} {\bibfield  {journal} {\bibinfo  {journal}
  {Monthly Notices of the Royal Astronomical Society}\ }\textbf {\bibinfo
  {volume} {370}},\ \bibinfo {pages} {1008} (\bibinfo {year}
  {2006}{\natexlab{b}})}\BibitemShut {NoStop}%
\bibitem [{\citenamefont {{Parker}}\ \emph {et~al.}(2007)\citenamefont
  {{Parker}}, \citenamefont {{Hoekstra}}, \citenamefont {{Hudson}},
  \citenamefont {{van Waerbeke}},\ and\ \citenamefont
  {{Mellier}}}]{Parker2007}%
  \BibitemOpen
  \bibfield  {author} {\bibinfo {author} {\bibfnamefont {L.~C.}\ \bibnamefont
  {{Parker}}}, \bibinfo {author} {\bibfnamefont {H.}~\bibnamefont
  {{Hoekstra}}}, \bibinfo {author} {\bibfnamefont {M.~J.}\ \bibnamefont
  {{Hudson}}}, \bibinfo {author} {\bibfnamefont {L.}~\bibnamefont {{van
  Waerbeke}}}, \ and\ \bibinfo {author} {\bibfnamefont {Y.}~\bibnamefont
  {{Mellier}}},\ }\href@noop {} {\bibfield  {journal} {\bibinfo  {journal}
  {\apj}\ }\textbf {\bibinfo {volume} {669}},\ \bibinfo {pages} {21} (\bibinfo
  {year} {2007})}\BibitemShut {NoStop}%
\bibitem [{\citenamefont {{van Uitert}}\ \emph {et~al.}(2012)\citenamefont
  {{van Uitert}}, \citenamefont {Hoekstra}, \citenamefont {Schrabback},
  \citenamefont {Gilbank}, \citenamefont {Gladders},\ and\ \citenamefont
  {Yee}}]{Uitert}%
  \BibitemOpen
  \bibfield  {author} {\bibinfo {author} {\bibfnamefont {E.}~\bibnamefont {{van
  Uitert}}}, \bibinfo {author} {\bibfnamefont {H.}~\bibnamefont {Hoekstra}},
  \bibinfo {author} {\bibfnamefont {T.}~\bibnamefont {Schrabback}}, \bibinfo
  {author} {\bibfnamefont {D.~G.}\ \bibnamefont {Gilbank}}, \bibinfo {author}
  {\bibfnamefont {M.~D.}\ \bibnamefont {Gladders}}, \ and\ \bibinfo {author}
  {\bibfnamefont {H.~K.~C.}\ \bibnamefont {Yee}},\ }\href@noop {} {\bibfield
  {journal} {\bibinfo  {journal} {Astronomy and Astrophysics}\ }\textbf
  {\bibinfo {volume} {545}} (\bibinfo {year} {2012})}\BibitemShut {NoStop}%
\bibitem [{\citenamefont {{Tenneti}}\ \emph {et~al.}(2014)\citenamefont
  {{Tenneti}}, \citenamefont {{Mandelbaum}}, \citenamefont {{Di Matteo}},
  \citenamefont {{Feng}},\ and\ \citenamefont {{Khandai}}}]{Tenneti2014}%
  \BibitemOpen
  \bibfield  {author} {\bibinfo {author} {\bibfnamefont {A.}~\bibnamefont
  {{Tenneti}}}, \bibinfo {author} {\bibfnamefont {R.}~\bibnamefont
  {{Mandelbaum}}}, \bibinfo {author} {\bibfnamefont {T.}~\bibnamefont {{Di
  Matteo}}}, \bibinfo {author} {\bibfnamefont {Y.}~\bibnamefont {{Feng}}}, \
  and\ \bibinfo {author} {\bibfnamefont {N.}~\bibnamefont {{Khandai}}},\ }\href
  {\doibase 10.1093/mnras/stu586} {\bibfield  {journal} {\bibinfo  {journal}
  {\mnras}\ }\textbf {\bibinfo {volume} {441}},\ \bibinfo {pages} {470}
  (\bibinfo {year} {2014})},\ \Eprint {http://arxiv.org/abs/1403.4215}
  {arXiv:1403.4215} \BibitemShut {NoStop}%
\bibitem [{\citenamefont {{Bett}}(2012)}]{Bett2012}%
  \BibitemOpen
  \bibfield  {author} {\bibinfo {author} {\bibfnamefont {P.}~\bibnamefont
  {{Bett}}},\ }\href {\doibase 10.1111/j.1365-2966.2011.20258.x} {\bibfield
  {journal} {\bibinfo  {journal} {\mnras}\ }\textbf {\bibinfo {volume} {420}},\
  \bibinfo {pages} {3303} (\bibinfo {year} {2012})},\ \Eprint
  {http://arxiv.org/abs/1108.3717} {arXiv:1108.3717 [astro-ph.CO]} \BibitemShut
  {NoStop}%
\bibitem [{\citenamefont {{Howell}}\ and\ \citenamefont
  {{Brainerd}}(2010)}]{Howell2010}%
  \BibitemOpen
  \bibfield  {author} {\bibinfo {author} {\bibfnamefont {P.~J.}\ \bibnamefont
  {{Howell}}}\ and\ \bibinfo {author} {\bibfnamefont {T.~G.}\ \bibnamefont
  {{Brainerd}}},\ }\href {\doibase 10.1111/j.1365-2966.2010.16979.x} {\bibfield
   {journal} {\bibinfo  {journal} {\mnras}\ }\textbf {\bibinfo {volume}
  {407}},\ \bibinfo {pages} {891} (\bibinfo {year} {2010})},\ \Eprint
  {http://arxiv.org/abs/1004.2491} {arXiv:1004.2491 [astro-ph.CO]} \BibitemShut
  {NoStop}%
\bibitem [{\citenamefont {Simon}\ \emph {et~al.}(2012)\citenamefont {Simon},
  \citenamefont {Schneider},\ and\ \citenamefont {K{\"u}bler}}]{Simon2012}%
  \BibitemOpen
  \bibfield  {author} {\bibinfo {author} {\bibfnamefont {P.}~\bibnamefont
  {Simon}}, \bibinfo {author} {\bibfnamefont {P.}~\bibnamefont {Schneider}}, \
  and\ \bibinfo {author} {\bibfnamefont {D.}~\bibnamefont {K{\"u}bler}},\
  }\href@noop {} {\bibfield  {journal} {\bibinfo  {journal} {Astronomy \&
  Astrophysics/Astronomie et Astrophysique}\ }\textbf {\bibinfo {volume} {548}}
  (\bibinfo {year} {2012})}\BibitemShut {NoStop}%
\bibitem [{\citenamefont {{Simon}}\ \emph {et~al.}(2013)\citenamefont
  {{Simon}}, \citenamefont {{Erben}}, \citenamefont {{Schneider}},
  \citenamefont {{Heymans}}, \citenamefont {{Hildebrandt}}, \citenamefont
  {{Hoekstra}}, \citenamefont {{Kitching}}, \citenamefont {{Mellier}},
  \citenamefont {{Miller}}, \citenamefont {{Van Waerbeke}}, \citenamefont
  {{Bonnett}}, \citenamefont {{Coupon}}, \citenamefont {{Fu}}, \citenamefont
  {{Hudson}}, \citenamefont {{Kuijken}}, \citenamefont {{Rowe}}, \citenamefont
  {{Schrabback}}, \citenamefont {{Semboloni}},\ and\ \citenamefont
  {{Velander}}}]{Simon2013}%
  \BibitemOpen
  \bibfield  {author} {\bibinfo {author} {\bibfnamefont {P.}~\bibnamefont
  {{Simon}}}, \bibinfo {author} {\bibfnamefont {T.}~\bibnamefont {{Erben}}},
  \bibinfo {author} {\bibfnamefont {P.}~\bibnamefont {{Schneider}}}, \bibinfo
  {author} {\bibfnamefont {C.}~\bibnamefont {{Heymans}}}, \bibinfo {author}
  {\bibfnamefont {H.}~\bibnamefont {{Hildebrandt}}}, \bibinfo {author}
  {\bibfnamefont {H.}~\bibnamefont {{Hoekstra}}}, \bibinfo {author}
  {\bibfnamefont {T.~D.}\ \bibnamefont {{Kitching}}}, \bibinfo {author}
  {\bibfnamefont {Y.}~\bibnamefont {{Mellier}}}, \bibinfo {author}
  {\bibfnamefont {L.}~\bibnamefont {{Miller}}}, \bibinfo {author}
  {\bibfnamefont {L.}~\bibnamefont {{Van Waerbeke}}}, \bibinfo {author}
  {\bibfnamefont {C.}~\bibnamefont {{Bonnett}}}, \bibinfo {author}
  {\bibfnamefont {J.}~\bibnamefont {{Coupon}}}, \bibinfo {author}
  {\bibfnamefont {L.}~\bibnamefont {{Fu}}}, \bibinfo {author} {\bibfnamefont
  {M.~J.}\ \bibnamefont {{Hudson}}}, \bibinfo {author} {\bibfnamefont
  {K.}~\bibnamefont {{Kuijken}}}, \bibinfo {author} {\bibfnamefont {B.~T.~P.}\
  \bibnamefont {{Rowe}}}, \bibinfo {author} {\bibfnamefont {T.}~\bibnamefont
  {{Schrabback}}}, \bibinfo {author} {\bibfnamefont {E.}~\bibnamefont
  {{Semboloni}}}, \ and\ \bibinfo {author} {\bibfnamefont {M.}~\bibnamefont
  {{Velander}}},\ }\href {\doibase 10.1093/mnras/stt069} {\bibfield  {journal}
  {\bibinfo  {journal} {\mnras}\ }\textbf {\bibinfo {volume} {430}},\ \bibinfo
  {pages} {2476} (\bibinfo {year} {2013})},\ \Eprint
  {http://arxiv.org/abs/1301.1863} {arXiv:1301.1863 [astro-ph.CO]} \BibitemShut
  {NoStop}%
\bibitem [{\citenamefont {{Fu}}\ \emph {et~al.}(2014)\citenamefont {{Fu}},
  \citenamefont {{Kilbinger}}, \citenamefont {{Erben}}, \citenamefont
  {{Heymans}}, \citenamefont {{Hildebrandt}}, \citenamefont {{Hoekstra}},
  \citenamefont {{Kitching}}, \citenamefont {{Mellier}}, \citenamefont
  {{Miller}}, \citenamefont {{Semboloni}}, \citenamefont {{Simon}},
  \citenamefont {{Van Waerbeke}}, \citenamefont {{Coupon}}, \citenamefont
  {{Harnois-D{\'e}raps}}, \citenamefont {{Hudson}}, \citenamefont {{Kuijken}},
  \citenamefont {{Rowe}}, \citenamefont {{Schrabback}}, \citenamefont
  {{Vafaei}},\ and\ \citenamefont {{Velander}}}]{Fu2014}%
  \BibitemOpen
  \bibfield  {author} {\bibinfo {author} {\bibfnamefont {L.}~\bibnamefont
  {{Fu}}}, \bibinfo {author} {\bibfnamefont {M.}~\bibnamefont {{Kilbinger}}},
  \bibinfo {author} {\bibfnamefont {T.}~\bibnamefont {{Erben}}}, \bibinfo
  {author} {\bibfnamefont {C.}~\bibnamefont {{Heymans}}}, \bibinfo {author}
  {\bibfnamefont {H.}~\bibnamefont {{Hildebrandt}}}, \bibinfo {author}
  {\bibfnamefont {H.}~\bibnamefont {{Hoekstra}}}, \bibinfo {author}
  {\bibfnamefont {T.~D.}\ \bibnamefont {{Kitching}}}, \bibinfo {author}
  {\bibfnamefont {Y.}~\bibnamefont {{Mellier}}}, \bibinfo {author}
  {\bibfnamefont {L.}~\bibnamefont {{Miller}}}, \bibinfo {author}
  {\bibfnamefont {E.}~\bibnamefont {{Semboloni}}}, \bibinfo {author}
  {\bibfnamefont {P.}~\bibnamefont {{Simon}}}, \bibinfo {author} {\bibfnamefont
  {L.}~\bibnamefont {{Van Waerbeke}}}, \bibinfo {author} {\bibfnamefont
  {J.}~\bibnamefont {{Coupon}}}, \bibinfo {author} {\bibfnamefont
  {J.}~\bibnamefont {{Harnois-D{\'e}raps}}}, \bibinfo {author} {\bibfnamefont
  {M.~J.}\ \bibnamefont {{Hudson}}}, \bibinfo {author} {\bibfnamefont
  {K.}~\bibnamefont {{Kuijken}}}, \bibinfo {author} {\bibfnamefont
  {B.}~\bibnamefont {{Rowe}}}, \bibinfo {author} {\bibfnamefont
  {T.}~\bibnamefont {{Schrabback}}}, \bibinfo {author} {\bibfnamefont
  {S.}~\bibnamefont {{Vafaei}}}, \ and\ \bibinfo {author} {\bibfnamefont
  {M.}~\bibnamefont {{Velander}}},\ }\href {\doibase 10.1093/mnras/stu754}
  {\bibfield  {journal} {\bibinfo  {journal} {\mnras}\ }\textbf {\bibinfo
  {volume} {441}},\ \bibinfo {pages} {2725} (\bibinfo {year} {2014})},\ \Eprint
  {http://arxiv.org/abs/1404.5469} {arXiv:1404.5469} \BibitemShut {NoStop}%
\bibitem [{\citenamefont {{Klypin}}\ \emph {et~al.}(2011)\citenamefont
  {{Klypin}}, \citenamefont {{Trujillo-Gomez}},\ and\ \citenamefont
  {{Primack}}}]{Klypin2011}%
  \BibitemOpen
  \bibfield  {author} {\bibinfo {author} {\bibfnamefont {A.~A.}\ \bibnamefont
  {{Klypin}}}, \bibinfo {author} {\bibfnamefont {S.}~\bibnamefont
  {{Trujillo-Gomez}}}, \ and\ \bibinfo {author} {\bibfnamefont
  {J.}~\bibnamefont {{Primack}}},\ }\href {\doibase
  10.1088/0004-637X/740/2/102} {\bibfield  {journal} {\bibinfo  {journal}
  {\apj}\ }\textbf {\bibinfo {volume} {740}},\ \bibinfo {eid} {102} (\bibinfo
  {year} {2011})},\ \Eprint {http://arxiv.org/abs/1002.3660} {arXiv:1002.3660
  [astro-ph.CO]} \BibitemShut {NoStop}%
\bibitem [{\citenamefont {Bartelmann}\ and\ \citenamefont
  {Schneider}(2001)}]{bartelmann2001weak}%
  \BibitemOpen
  \bibfield  {author} {\bibinfo {author} {\bibfnamefont {M.}~\bibnamefont
  {Bartelmann}}\ and\ \bibinfo {author} {\bibfnamefont {P.}~\bibnamefont
  {Schneider}},\ }\href@noop {} {\bibfield  {journal} {\bibinfo  {journal}
  {Physics Reports}\ }\textbf {\bibinfo {volume} {340}},\ \bibinfo {pages}
  {291} (\bibinfo {year} {2001})}\BibitemShut {NoStop}%
\bibitem [{\citenamefont {{Diemer}}\ and\ \citenamefont
  {{Kravtsov}}(2014)}]{diemer2014dependence}%
  \BibitemOpen
  \bibfield  {author} {\bibinfo {author} {\bibfnamefont {B.}~\bibnamefont
  {{Diemer}}}\ and\ \bibinfo {author} {\bibfnamefont {A.~V.}\ \bibnamefont
  {{Kravtsov}}},\ }\href@noop {} {\bibfield  {journal} {\bibinfo  {journal}
  {arXiv preprint arXiv:1401.1216}\ } (\bibinfo {year} {2014})}\BibitemShut
  {NoStop}%
\bibitem [{Note1()}]{Note1}%
  \BibitemOpen
  \bibinfo {note} {Http://hipacc.ucsc.edu/Bolshoi/MergerTrees.html}\BibitemShut
  {NoStop}%
\bibitem [{\citenamefont {Peebles}(1993)}]{peebles}%
  \BibitemOpen
  \bibfield  {author} {\bibinfo {author} {\bibfnamefont {P.~J.~E.}\
  \bibnamefont {Peebles}},\ }\href@noop {} {\emph {\bibinfo {title} {Principles
  of physical cosmology}}}\ (\bibinfo  {publisher} {Princeton University
  Press},\ \bibinfo {year} {1993})\BibitemShut {NoStop}%
\bibitem [{\citenamefont {Chang}\ \emph {et~al.}(2013)\citenamefont {Chang},
  \citenamefont {Jarvis}, \citenamefont {Jain}, \citenamefont {Kahn},
  \citenamefont {Kirkby}, \citenamefont {Connolly}, \citenamefont {Krughoff},
  \citenamefont {Peng},\ and\ \citenamefont {Peterson}}]{Chang}%
  \BibitemOpen
  \bibfield  {author} {\bibinfo {author} {\bibfnamefont {C.}~\bibnamefont
  {Chang}}, \bibinfo {author} {\bibfnamefont {M.}~\bibnamefont {Jarvis}},
  \bibinfo {author} {\bibfnamefont {B.}~\bibnamefont {Jain}}, \bibinfo {author}
  {\bibfnamefont {S.}~\bibnamefont {Kahn}}, \bibinfo {author} {\bibfnamefont
  {D.}~\bibnamefont {Kirkby}}, \bibinfo {author} {\bibfnamefont
  {A.}~\bibnamefont {Connolly}}, \bibinfo {author} {\bibfnamefont
  {S.}~\bibnamefont {Krughoff}}, \bibinfo {author} {\bibfnamefont {E.-H.}\
  \bibnamefont {Peng}}, \ and\ \bibinfo {author} {\bibfnamefont
  {J.}~\bibnamefont {Peterson}},\ }\href@noop {} {\bibfield  {journal}
  {\bibinfo  {journal} {Monthly Notices of the Royal Astronomical Society}\
  }\textbf {\bibinfo {volume} {434}},\ \bibinfo {pages} {2121} (\bibinfo {year}
  {2013})}\BibitemShut {NoStop}%
\bibitem [{\citenamefont {{Croft}}\ and\ \citenamefont
  {{Metzler}}(2000)}]{2000ApJ...545..561C}%
  \BibitemOpen
  \bibfield  {author} {\bibinfo {author} {\bibfnamefont {R.~A.~C.}\
  \bibnamefont {{Croft}}}\ and\ \bibinfo {author} {\bibfnamefont {C.~A.}\
  \bibnamefont {{Metzler}}},\ }\href {\doibase 10.1086/317856} {\bibfield
  {journal} {\bibinfo  {journal} {\apj}\ }\textbf {\bibinfo {volume} {545}},\
  \bibinfo {pages} {561} (\bibinfo {year} {2000})},\ \Eprint
  {http://arxiv.org/abs/astro-ph/0005384} {astro-ph/0005384} \BibitemShut
  {NoStop}%
\bibitem [{\citenamefont {{Heavens}}\ \emph {et~al.}(2000)\citenamefont
  {{Heavens}}, \citenamefont {{Refregier}},\ and\ \citenamefont
  {{Heymans}}}]{2000MNRAS.319..649H}%
  \BibitemOpen
  \bibfield  {author} {\bibinfo {author} {\bibfnamefont {A.}~\bibnamefont
  {{Heavens}}}, \bibinfo {author} {\bibfnamefont {A.}~\bibnamefont
  {{Refregier}}}, \ and\ \bibinfo {author} {\bibfnamefont {C.}~\bibnamefont
  {{Heymans}}},\ }\href {\doibase 10.1046/j.1365-8711.2000.03907.x} {\bibfield
  {journal} {\bibinfo  {journal} {\mnras}\ }\textbf {\bibinfo {volume} {319}},\
  \bibinfo {pages} {649} (\bibinfo {year} {2000})},\ \Eprint
  {http://arxiv.org/abs/astro-ph/0005269} {astro-ph/0005269} \BibitemShut
  {NoStop}%
\bibitem [{\citenamefont {{Crittenden}}\ \emph {et~al.}(2001)\citenamefont
  {{Crittenden}}, \citenamefont {{Natarajan}}, \citenamefont {{Pen}},\ and\
  \citenamefont {{Theuns}}}]{2001ApJ...559..552C}%
  \BibitemOpen
  \bibfield  {author} {\bibinfo {author} {\bibfnamefont {R.~G.}\ \bibnamefont
  {{Crittenden}}}, \bibinfo {author} {\bibfnamefont {P.}~\bibnamefont
  {{Natarajan}}}, \bibinfo {author} {\bibfnamefont {U.-L.}\ \bibnamefont
  {{Pen}}}, \ and\ \bibinfo {author} {\bibfnamefont {T.}~\bibnamefont
  {{Theuns}}},\ }\href {\doibase 10.1086/322370} {\bibfield  {journal}
  {\bibinfo  {journal} {\apj}\ }\textbf {\bibinfo {volume} {559}},\ \bibinfo
  {pages} {552} (\bibinfo {year} {2001})},\ \Eprint
  {http://arxiv.org/abs/astro-ph/0009052} {astro-ph/0009052} \BibitemShut
  {NoStop}%
\bibitem [{\citenamefont {{Hirata}}\ and\ \citenamefont
  {{Seljak}}(2004)}]{2004PhRvD..70f3526H}%
  \BibitemOpen
  \bibfield  {author} {\bibinfo {author} {\bibfnamefont {C.~M.}\ \bibnamefont
  {{Hirata}}}\ and\ \bibinfo {author} {\bibfnamefont {U.}~\bibnamefont
  {{Seljak}}},\ }\href {\doibase 10.1103/PhysRevD.70.063526} {\bibfield
  {journal} {\bibinfo  {journal} {\prd}\ }\textbf {\bibinfo {volume} {70}},\
  \bibinfo {eid} {063526} (\bibinfo {year} {2004})},\ \Eprint
  {http://arxiv.org/abs/astro-ph/0406275} {astro-ph/0406275} \BibitemShut
  {NoStop}%
\bibitem [{\citenamefont {{Hirata}}\ \emph {et~al.}(2004)\citenamefont
  {{Hirata}}, \citenamefont {{Mandelbaum}}, \citenamefont {{Seljak}},
  \citenamefont {{Guzik}}, \citenamefont {{Padmanabhan}}, \citenamefont
  {{Blake}}, \citenamefont {{Brinkmann}}, \citenamefont {{Bud{\'a}vari}},
  \citenamefont {{Connolly}}, \citenamefont {{Csabai}}, \citenamefont
  {{Scranton}},\ and\ \citenamefont {{Szalay}}}]{2004MNRAS.353..529H}%
  \BibitemOpen
  \bibfield  {author} {\bibinfo {author} {\bibfnamefont {C.~M.}\ \bibnamefont
  {{Hirata}}}, \bibinfo {author} {\bibfnamefont {R.}~\bibnamefont
  {{Mandelbaum}}}, \bibinfo {author} {\bibfnamefont {U.}~\bibnamefont
  {{Seljak}}}, \bibinfo {author} {\bibfnamefont {J.}~\bibnamefont {{Guzik}}},
  \bibinfo {author} {\bibfnamefont {N.}~\bibnamefont {{Padmanabhan}}}, \bibinfo
  {author} {\bibfnamefont {C.}~\bibnamefont {{Blake}}}, \bibinfo {author}
  {\bibfnamefont {J.}~\bibnamefont {{Brinkmann}}}, \bibinfo {author}
  {\bibfnamefont {T.}~\bibnamefont {{Bud{\'a}vari}}}, \bibinfo {author}
  {\bibfnamefont {A.}~\bibnamefont {{Connolly}}}, \bibinfo {author}
  {\bibfnamefont {I.}~\bibnamefont {{Csabai}}}, \bibinfo {author}
  {\bibfnamefont {R.}~\bibnamefont {{Scranton}}}, \ and\ \bibinfo {author}
  {\bibfnamefont {A.~S.}\ \bibnamefont {{Szalay}}},\ }\href {\doibase
  10.1111/j.1365-2966.2004.08090.x} {\bibfield  {journal} {\bibinfo  {journal}
  {\mnras}\ }\textbf {\bibinfo {volume} {353}},\ \bibinfo {pages} {529}
  (\bibinfo {year} {2004})},\ \Eprint {http://arxiv.org/abs/astro-ph/0403255}
  {astro-ph/0403255} \BibitemShut {NoStop}%
\bibitem [{\citenamefont {{Mandelbaum}}\ \emph
  {et~al.}(2006{\natexlab{c}})\citenamefont {{Mandelbaum}}, \citenamefont
  {{Hirata}}, \citenamefont {{Ishak}}, \citenamefont {{Seljak}},\ and\
  \citenamefont {{Brinkmann}}}]{2006MNRAS.367..611M}%
  \BibitemOpen
  \bibfield  {author} {\bibinfo {author} {\bibfnamefont {R.}~\bibnamefont
  {{Mandelbaum}}}, \bibinfo {author} {\bibfnamefont {C.~M.}\ \bibnamefont
  {{Hirata}}}, \bibinfo {author} {\bibfnamefont {M.}~\bibnamefont {{Ishak}}},
  \bibinfo {author} {\bibfnamefont {U.}~\bibnamefont {{Seljak}}}, \ and\
  \bibinfo {author} {\bibfnamefont {J.}~\bibnamefont {{Brinkmann}}},\ }\href
  {\doibase 10.1111/j.1365-2966.2005.09946.x} {\bibfield  {journal} {\bibinfo
  {journal} {\mnras}\ }\textbf {\bibinfo {volume} {367}},\ \bibinfo {pages}
  {611} (\bibinfo {year} {2006}{\natexlab{c}})},\ \Eprint
  {http://arxiv.org/abs/astro-ph/0509026} {astro-ph/0509026} \BibitemShut
  {NoStop}%
\bibitem [{\citenamefont {{Hirata}}\ \emph {et~al.}(2007)\citenamefont
  {{Hirata}}, \citenamefont {{Mandelbaum}}, \citenamefont {{Ishak}},
  \citenamefont {{Seljak}}, \citenamefont {{Nichol}}, \citenamefont
  {{Pimbblet}}, \citenamefont {{Ross}},\ and\ \citenamefont
  {{Wake}}}]{2007MNRAS.381.1197H}%
  \BibitemOpen
  \bibfield  {author} {\bibinfo {author} {\bibfnamefont {C.~M.}\ \bibnamefont
  {{Hirata}}}, \bibinfo {author} {\bibfnamefont {R.}~\bibnamefont
  {{Mandelbaum}}}, \bibinfo {author} {\bibfnamefont {M.}~\bibnamefont
  {{Ishak}}}, \bibinfo {author} {\bibfnamefont {U.}~\bibnamefont {{Seljak}}},
  \bibinfo {author} {\bibfnamefont {R.}~\bibnamefont {{Nichol}}}, \bibinfo
  {author} {\bibfnamefont {K.~A.}\ \bibnamefont {{Pimbblet}}}, \bibinfo
  {author} {\bibfnamefont {N.~P.}\ \bibnamefont {{Ross}}}, \ and\ \bibinfo
  {author} {\bibfnamefont {D.}~\bibnamefont {{Wake}}},\ }\href {\doibase
  10.1111/j.1365-2966.2007.12312.x} {\bibfield  {journal} {\bibinfo  {journal}
  {\mnras}\ }\textbf {\bibinfo {volume} {381}},\ \bibinfo {pages} {1197}
  (\bibinfo {year} {2007})},\ \Eprint {http://arxiv.org/abs/astro-ph/0701671}
  {astro-ph/0701671} \BibitemShut {NoStop}%
\bibitem [{\citenamefont {{Blazek}}\ \emph {et~al.}(2011)\citenamefont
  {{Blazek}}, \citenamefont {{McQuinn}},\ and\ \citenamefont
  {{Seljak}}}]{Blazek2011}%
  \BibitemOpen
  \bibfield  {author} {\bibinfo {author} {\bibfnamefont {J.}~\bibnamefont
  {{Blazek}}}, \bibinfo {author} {\bibfnamefont {M.}~\bibnamefont {{McQuinn}}},
  \ and\ \bibinfo {author} {\bibfnamefont {U.}~\bibnamefont {{Seljak}}},\
  }\href {\doibase 10.1088/1475-7516/2011/05/010} {\bibfield  {journal}
  {\bibinfo  {journal} {\jcap}\ }\textbf {\bibinfo {volume} {5}},\ \bibinfo
  {eid} {010} (\bibinfo {year} {2011})},\ \Eprint
  {http://arxiv.org/abs/1101.4017} {arXiv:1101.4017 [astro-ph.CO]} \BibitemShut
  {NoStop}%
\bibitem [{\citenamefont {{Blazek}}\ \emph {et~al.}(2012)\citenamefont
  {{Blazek}}, \citenamefont {{Mandelbaum}}, \citenamefont {{Seljak}},\ and\
  \citenamefont {{Nakajima}}}]{Blazek2012}%
  \BibitemOpen
  \bibfield  {author} {\bibinfo {author} {\bibfnamefont {J.}~\bibnamefont
  {{Blazek}}}, \bibinfo {author} {\bibfnamefont {R.}~\bibnamefont
  {{Mandelbaum}}}, \bibinfo {author} {\bibfnamefont {U.}~\bibnamefont
  {{Seljak}}}, \ and\ \bibinfo {author} {\bibfnamefont {R.}~\bibnamefont
  {{Nakajima}}},\ }\href {\doibase 10.1088/1475-7516/2012/05/041} {\bibfield
  {journal} {\bibinfo  {journal} {\jcap}\ }\textbf {\bibinfo {volume} {5}},\
  \bibinfo {eid} {041} (\bibinfo {year} {2012})},\ \Eprint
  {http://arxiv.org/abs/1204.2264} {arXiv:1204.2264 [astro-ph.CO]} \BibitemShut
  {NoStop}%
\bibitem [{\citenamefont {{Jarvis}}\ \emph {et~al.}(2004)\citenamefont
  {{Jarvis}}, \citenamefont {{Takada}}, \citenamefont {{Jain}},\ and\
  \citenamefont {{Bernstein}}}]{Jarvis2004}%
  \BibitemOpen
  \bibfield  {author} {\bibinfo {author} {\bibfnamefont {M.}~\bibnamefont
  {{Jarvis}}}, \bibinfo {author} {\bibfnamefont {M.}~\bibnamefont {{Takada}}},
  \bibinfo {author} {\bibfnamefont {B.}~\bibnamefont {{Jain}}}, \ and\ \bibinfo
  {author} {\bibfnamefont {G.}~\bibnamefont {{Bernstein}}},\ }in\ \href@noop {}
  {\emph {\bibinfo {booktitle} {American Astronomical Society Meeting
  Abstracts}}},\ \bibinfo {series} {Bulletin of the American Astronomical
  Society}, Vol.~\bibinfo {volume} {36}\ (\bibinfo {year} {2004})\ p.\ \bibinfo
  {pages} {108.17}\BibitemShut {NoStop}%
\bibitem [{\citenamefont {{Semboloni}}\ \emph {et~al.}(2011)\citenamefont
  {{Semboloni}}, \citenamefont {{Schrabback}}, \citenamefont {{van Waerbeke}},
  \citenamefont {{Vafaei}}, \citenamefont {{Hartlap}},\ and\ \citenamefont
  {{Hilbert}}}]{Semboloni2011}%
  \BibitemOpen
  \bibfield  {author} {\bibinfo {author} {\bibfnamefont {E.}~\bibnamefont
  {{Semboloni}}}, \bibinfo {author} {\bibfnamefont {T.}~\bibnamefont
  {{Schrabback}}}, \bibinfo {author} {\bibfnamefont {L.}~\bibnamefont {{van
  Waerbeke}}}, \bibinfo {author} {\bibfnamefont {S.}~\bibnamefont {{Vafaei}}},
  \bibinfo {author} {\bibfnamefont {J.}~\bibnamefont {{Hartlap}}}, \ and\
  \bibinfo {author} {\bibfnamefont {S.}~\bibnamefont {{Hilbert}}},\ }\href
  {\doibase 10.1111/j.1365-2966.2010.17430.x} {\bibfield  {journal} {\bibinfo
  {journal} {\mnras}\ }\textbf {\bibinfo {volume} {410}},\ \bibinfo {pages}
  {143} (\bibinfo {year} {2011})},\ \Eprint {http://arxiv.org/abs/1005.4941}
  {arXiv:1005.4941 [astro-ph.CO]} \BibitemShut {NoStop}%
\bibitem [{\citenamefont {{Kazantzidis}}\ \emph {et~al.}(2004)\citenamefont
  {{Kazantzidis}}, \citenamefont {{Kravtsov}}, \citenamefont {{Zentner}},
  \citenamefont {{Allgood}}, \citenamefont {{Nagai}},\ and\ \citenamefont
  {{Moore}}}]{Kazantzidis2004}%
  \BibitemOpen
  \bibfield  {author} {\bibinfo {author} {\bibfnamefont {S.}~\bibnamefont
  {{Kazantzidis}}}, \bibinfo {author} {\bibfnamefont {A.~V.}\ \bibnamefont
  {{Kravtsov}}}, \bibinfo {author} {\bibfnamefont {A.~R.}\ \bibnamefont
  {{Zentner}}}, \bibinfo {author} {\bibfnamefont {B.}~\bibnamefont
  {{Allgood}}}, \bibinfo {author} {\bibfnamefont {D.}~\bibnamefont {{Nagai}}},
  \ and\ \bibinfo {author} {\bibfnamefont {B.}~\bibnamefont {{Moore}}},\ }\href
  {\doibase 10.1086/423992} {\bibfield  {journal} {\bibinfo  {journal} {\apjl}\
  }\textbf {\bibinfo {volume} {611}},\ \bibinfo {pages} {L73} (\bibinfo {year}
  {2004})},\ \Eprint {http://arxiv.org/abs/astro-ph/0405189} {astro-ph/0405189}
  \BibitemShut {NoStop}%
\bibitem [{\citenamefont {{Bryan}}\ \emph {et~al.}(2013)\citenamefont
  {{Bryan}}, \citenamefont {{Kay}}, \citenamefont {{Duffy}}, \citenamefont
  {{Schaye}}, \citenamefont {{Dalla Vecchia}},\ and\ \citenamefont
  {{Booth}}}]{Bryan2013}%
  \BibitemOpen
  \bibfield  {author} {\bibinfo {author} {\bibfnamefont {S.~E.}\ \bibnamefont
  {{Bryan}}}, \bibinfo {author} {\bibfnamefont {S.~T.}\ \bibnamefont {{Kay}}},
  \bibinfo {author} {\bibfnamefont {A.~R.}\ \bibnamefont {{Duffy}}}, \bibinfo
  {author} {\bibfnamefont {J.}~\bibnamefont {{Schaye}}}, \bibinfo {author}
  {\bibfnamefont {C.}~\bibnamefont {{Dalla Vecchia}}}, \ and\ \bibinfo {author}
  {\bibfnamefont {C.~M.}\ \bibnamefont {{Booth}}},\ }\href {\doibase
  10.1093/mnras/sts587} {\bibfield  {journal} {\bibinfo  {journal} {\mnras}\
  }\textbf {\bibinfo {volume} {429}},\ \bibinfo {pages} {3316} (\bibinfo {year}
  {2013})},\ \Eprint {http://arxiv.org/abs/1207.4555} {arXiv:1207.4555
  [astro-ph.CO]} \BibitemShut {NoStop}%
\end{thebibliography}%

\end{document}